\documentclass[a4paper, 11pt]{article}
\RequirePackage[OT1]{fontenc}
\usepackage{amsfonts,amssymb,graphics,epsfig,verbatim,bm,latexsym,amsmath,url,float,amsbsy,multirow,multicol,rotating,enumerate,graphicx,pgf,tikz,subfig,color,xcolor,lscape}
\usepackage{amsthm,mathtools}
\usepackage[textheight=722pt,textwidth=500pt,centering,headheight=50pt,headsep=12pt,footskip=12pt]{geometry}
\usepackage[colorlinks,citecolor=blue,linkcolor=purple!60!black,urlcolor=purple!60!black]{hyperref}
\usepackage[framemethod=TikZ]{mdframed}
\usepackage[numbers,sort&compress]{natbib}
\usepackage{booktabs} 
\usetikzlibrary{positioning}
\usetikzlibrary[shapes.misc]
\usetikzlibrary{arrows,automata}
\usetikzlibrary{decorations.pathreplacing}
\allowdisplaybreaks
\makeatletter
\let\@fnsymbol\@arabic
\makeatother

\newcommand{\indep}{\perp \!\!\! \perp}

\newtheorem{corollary}{Corollary}%

\newtheorem{definition}{Definition}

\title{\vspace*{-2cm}\LARGE\bf Recoverability of Causal Effects under Presence of Missing Data: a Longitudinal Case Study}
\date{}
\author{Anastasiia Holovchak\footnote{Seminar for Statistics, ETH Zurich, Switzerland, \href{anastasiia.holovchak@stat.math.ethz.ch}{anastasiia.holovchak@stat.math.ethz.ch}}
       \and
       Helen McIlleron\footnote{Division of Clinical Pharmacology, Department of Medicine, University of Cape Town, South Africa}
       \and
       Paolo Denti$^{\text{2}}$
       \and
       Michael Schomaker\footnote{Department of Statistics, Ludwig-Maximilians University Munich, Germany}~$^{,}$\footnote{Centre for Infectious Disease Epidemiology and Research, University of Cape Town, South Africa}
       }
\begin{document}

\maketitle

\begin{abstract}
Missing data in multiple variables is a common issue. We investigate the applicability of the framework of graphical models for handling missing data to a complex longitudinal pharmacological study of children with HIV treated with an efavirenz-based regimen as part of the CHAPAS-3 trial. Specifically, we examine whether the causal effects of interest, defined through static interventions on multiple continuous variables, can be recovered (estimated consistently) from the available data only. So far, no general algorithms are available to decide on recoverability, and decisions have to be made on a case-by-case basis. We emphasize sensitivity of recoverability to even the smallest changes in the graph structure, and present recoverability results for three plausible missingness directed acyclic graphs (m-DAGs) in the CHAPAS-3 study, informed by clinical knowledge. Furthermore, we propose the concept of ``closed missingness mechanisms'' and show that under these mechanisms an available case analysis is admissible for consistent estimation for any type of statistical and causal query, even if the underlying missingness mechanism is of missing not at random (MNAR) type. Both simulations and theoretical considerations demonstrate how, in the assumed MNAR setting of our study, a complete or available case analysis can be superior to multiple imputation, and estimation results vary depending on the assumed missingness DAG. Our analyses demonstrate an innovative application of missingness DAGs to complex longitudinal real-world data, while highlighting the sensitivity of the results with respect to the assumed causal model.
\end{abstract}

\section{Introduction}\label{sec1}

Missing data are a common issue in biomedical research. In particular, longitudinal epidemiological studies, where data are collected consecutively at several points in time, tend to suffer from missing data on multiple variables. One common assumption which is often made regarding the missing data mechanism is the so-called \emph{missing at random} (MAR) assumption. If this assumption holds, statistical methods for consistent estimation are available, for example, multiple imputation or inverse probability weighting \citep{rubin1976,mohanNIPS2014}.

However, if multiple variables are missing simultaneously, it is often difficult to justify whether the MAR assumption holds. This is both because the classic MAR definition is event- rather than variable-based, and arguing for or against conditional independencies of \emph{events} is practically challenging, especially in longitudinal settings \citep{mohan2021}. Moreover, even though MAR is the weakest known condition under which the missingness process can be ignored (i.e., dealt with using the observed data), it is only a sufficient, but not a necessary condition for unbiased estimation; this means that under a \emph{missing not at random} (MNAR) scenario, it is unclear if and how a target estimand can be recovered (estimated consistently).

To tackle these and other challenges, \citet{mohan2013} and \citet{mohan2021} proposed an alternative causal graph-based framework, in which knowledge and assumptions about \emph{reasons} for missingness are encoded in relationships between \emph{variables} and missing data indicators. This framework is very general and can be used to evaluate whether estimands can be recovered given a correct causal missingness model (missingness directed acyclic graph, m-DAG).

However, currently there is no general algorithm that can be used to establish recoverability for arbitrary settings. Therefore, identification and estimation strategies have to be developed on a case-by-case basis for each particular scenario. This poses the question of how useful causal m-DAGs are practically.

\citet{moreno2018} have convincingly argued that one should thus develop canonical m-DAGs for recurring settings, such as for point-exposure study designs in epidemiology where missing data in outcome, exposure, and confounders are caused by some ``standard'' mechanisms. They demonstrated the usefulness of this approach in a study investigating the relationship between maternal mental illness and child behavior.

While causal missingness graphs, and their canonical versions, are a major advancement for causal inference research under missing data, their actual applicability has yet to be demonstrated in complex longitudinal studies. It is unclear how well knowledge on missing data can actually be  collected, then integrated in a realistic causal graph, and how difficult the mathematical exercise of establishing identification and recoverability results in such a complex, yet realistic setting is. Can m-DAGs make their way from blackboards to actual applications?

Recent work, such as by \citet{balzer2023} and \citet{nugent2023}, has advanced the field by using causal graphs and corresponding estimators to address recoverability and estimation of causal effects in longitudinal settings subject to complex missingness and dependence. These studies focus on reducing bias and improving efficiency by controlling for missing outcomes when estimating intervention effects. Their frameworks, both relying on a  Two-Stage TMLE approach, are particularly useful for handling data in complex cluster randomized trial settings. 

In this work, we focus on the explicit modeling of the missingness mechanism, not only for the outcome but for all relevant variables in the study. This involves developing a causal missingness model, explaining its motivation through a variable-based taxonomy, and then demonstrating whether identification is possible and what estimation strategies could be employed. We present an innovative application of missingness DAGs to a longitudinal study. Specifically, we investigate whether the derivation of identification results in a longitudinal setting is feasible, how volatile those results are, and how complex deriving those results can be.

Additionally, we conduct simulation studies aligned with the data example to, firstly, verify the theoretical results on the recoverability of the desired causal query, and secondly, to quantify the extent of bias in settings of special interest. Our simulations and theoretical considerations demonstrate that causal diagrams help to explicitly guide the process of decision-making about whether a parameter of interest can be consistently estimated from the available data. In practice, MAR is often assumed and multiple imputation is performed, which in many cases may lead to biased estimation results. On the other hand, we demonstrate that even in complex longitudinal study settings, there are MNAR scenarios for which available case analysis leads to valid estimation results, whereas multiple imputation does not.

We sought to answer these questions by applying the framework proposed by \citet{mohan2021} to answer a topical question in HIV-related pharmacoepidemiology, which is explained below in Section \ref{sec2}. We  i) investigate how clinical knowledge on reasons for our missing data can be best collected and integrated into a realistic  causal missingness graph, ii) derive recoverability results for our target estimand, iii) discuss estimation strategies for it, iv) evaluate the sensitivity of our results with respect to the assumed causal model, and v) seek for structures that may be helpful in establishing results for future studies.

This paper is structured as follows. We introduce the motivating question in Section \ref{sec2}, followed by the theoretical framework in Section \ref{sec3}. In Section \ref{sec4}, analyse the data from the complex longitudinal pharmacological study CHAPAS-3, followed by the simulation study in Section \ref{sec5}. We conclude in Section \ref{sec6}.

\section{Motivating Study}\label{sec2}

Our motivating data analysis comes from the Children with HIV in Africa–Pharmacokinetics and Adherence/Acceptability of Simple antiretroviral regimens randomized trial (CHAPAS-3). The study enrolled 478 children with HIV, between 1 month and 13 years of age, in 4 sites in Uganda and Zambia \citep{mulenga2016}. Children enrolled into the trial received combined antiretroviral therapy, i.e., one non-nucleoside reverse transcriptase inhibitor (efavirenz or nevirapine) and two nucleoside reverse transcriptase inhibitors (abacavir, stavudine or zidovudine --which were the randomized components-- and lamivudine).

We are interested in determining target concentrations using data from 125 children treated with efavirenz (EFV). Efavirenz is used not only in children, but also adults, though dosing recommendations (between 200 and 600 mg) depend on weight and age. Due to their different metabolic profiles and adherence patterns, patients with the same efavirenz dose may have different concentrations, conferring different protection against viral replication. Knowledge about concentrations is often used to derive dosing recommendations using population PK models \citep{bienczak2016}.  It is typically suggested that EFV concentrations between 1 and 4 mg/L should be achieved (at 12h after the dose was given, C12h). This is because lower concentrations may be insufficient to guarantee viral suppression and thus effective treatment, while higher concentrations may lead to toxicity negatively affecting the central nervous system. Our target estimand is thus the causal concentration-response curve (CCRC) at each follow-up visit, i.e. \emph{we are interested in how the counterfactual probability of viral failure at time $t$ varies as a function of possible prior concentration trajectories}.

Our analysis makes use of the data from \cite{bienczak2016}. We recently discussed statistical approaches on estimating the CCRC (\citealp{Schomaker:2023b}), but did not discuss aspects of missing data. In this manuscript, we extend the above study by developing a causal missingness graph (informed by the pediatrician's and trial team's knowledge) and derive identification and estimation approaches for our estimand of interest. More details on the analysis are given in Section \ref{sec4}.

\section{Framework}\label{sec3}

Missingness DAGs provide a natural extension of causal DAGs under the presence of missing data. Consider a DAG $\mathbf{G} = (\mathbf{V}, \mathbf{E})$ with a set of nodes $\mathbf{V}$, $|\mathbf{V}|=n$, which can be separated into two subsets $\mathbf{V}_o$ and $\mathbf{V}_m$, corresponding to sets of \emph{completely observed} and \emph{partially observed} variables. For each variable $X_i$, $i \in \{1, ..., n\}$, from the subset $\mathbf{V}_m$, a binary missingness indicator variable $M_{X_i}$
    $$M_{X_i} =
    \begin{cases}
        1 \quad \text{if $X_i$ is missing,} \\
        0 \quad \text{otherwise}
    \end{cases}$$
is introduced in the missingness DAG to depict causal relationships with other relevant variables. In the following, we refer to the set of missingness indicator variables as $M$. Additionally, the setup allows for the existence of a latent (unmeasured) variable set $\mathbf{U}$.

In this work, we make use of the framework introduced by \citet{mohan2021}. We refer to $\mathbf{G}_c$ as the \emph{complete-DAG (c-DAG)} (as in \citet{moreno2018}), and $\mathbf{G}$ as the \emph{missingness DAG (m-DAG)}. The c-DAG $\mathbf{G}_c$ includes only the variable set $\mathbf{V} := \mathbf{V}_o \cup \mathbf{V}_m \cup \mathbf{U}$ that is relevant for identification considerations with respect to the (causal) parameter of interest (assuming no missingness in the variables in $\mathbf{V}_m$). The m-DAG $\mathbf{G}$, however, also includes the set of missingness indicators $\mathbf{M}$, and may additionally consist of a set of auxiliary variables, denoted as $\mathbf{Z}$, with variables in $\mathbf{Z}$ causing missingness in the variables in $\mathbf{V}_m$, but not being a cause of the variables from the set $\mathbf{V}$. Note that variables  from both $\mathbf{V}$ and $\mathbf{Z}$ may be a cause of missingness in the variables from $\mathbf{V}_m$. \\

\cite{mohan2013}, \citet{mohan2021} show that causal diagrams can be used as a powerful tool for the identification and classification of missingness mechanisms. It is an important finding that the conventional taxonomy of missing data \citep{rubin1976} can be translated (with some minor changes) into the context of causal diagrams, which is the focus of this work. However, it is essential to mention that the missingness taxonomy definitions proposed by \citet{rubin1976} and \citet{mohan2013} are, in general, not equivalent. Note that Rubin's framework of missing data is defined on the \emph{record-based level}, which makes it particularly difficult to apply in practice. Another issue that is often ignored is that the `realized' MAR definition as proposed by \citet{rubin1976} is not required to hold for all possible values of the missingness indicator $M$, but only for those present in the data. This makes clear that such a kind of definition does not hold for a data-generating distribution in general, but only for a single data sample, as different missing data patterns may arise for samples resulting from re-running of the experiment \citep{schafer2002, tian2015}. To overcome this issue, \citet{seaman2013} distinguishes between two types of MAR - \emph{realized} MAR and \emph{everywhere} MAR. The latter is a `stronger' definition in the sense that conditional missingness distribution relates to the missing entries for all realized and unrealized patterns of missing data, and not only to the one specific data sample which has been observed.

\citet{mohan2013} propose an alternative framework of missing data taxonomy, which is based on graphical models. The two main advantages of the graph-based method are an explicit encoding of dependencies on the variable level (and not record-based), and also depiction of causal mechanisms which are causing missingness \citep{tian2015}. Note that m-DAGs represent both the data generating mechanism and the missing data mechanism, both in a causal manner.\\

Let $\mathbf{G}$ be an m-DAG over a set of variables $\mathbf{V}_o \cup \mathbf{V}_m \cup \mathbf{U} \cup \mathbf{M}$, where $\mathbf{V}_o$ and $\mathbf{V}_m$ denote the sets of completely and partially observed variables, correspondingly, $\mathbf{U}$ is a set of latent (unobserved) variables, and $\mathbf{M}$ is the set of missingness indicators. We denote the corresponding (joint) data distribution as $P(\mathbf{V}_o, \mathbf{V}_m, \mathbf{U}, \mathbf{M})$, and assume the distribution to be \emph{faithful} \citep{pearl1990equivalence} with respect to $\mathbf{G}$. Broadly speaking, faithfulness requires that the joint distribution $P(\mathbf{V}_o, \mathbf{V}_m, \mathbf{U}, \mathbf{M})$ satisfies all the conditional independence relationships encoded by the DAG, and only those relationships. In other words, one assumes that all observed conditional independencies follow from the graphical structure, and not from other reasons (such as deterministic relationships between variables). \\

The missingness mechanism is commonly characterized in terms of the conditional distribution of $\mathbf{M}$ given $(\mathbf{V}_o, \mathbf{V}_m, U)$, say $p(\mathbf{M}|\mathbf{V}_o, \mathbf{V}_m, U)$. It has to be assumed that any missingness indicator from the set $\mathbf{M}$ is not a parent of any variable from $\mathbf{V}_o \cup \mathbf{V}_m \cup U$. We emphasize once more that in the graph-based missing data framework, we work with variable-based and not the record-based definitions of the missingness mechanisms.

\begin{figure}[htb]
    \centering
    \subfloat[MCAR]{
    \scalebox{0.76}{
        \begin{tikzpicture}
            \node[color=black!40!gray] (m) at (-2.40,1.50) {M$_Y$};
            \node[fill=black!20!yellow] (v0) at (-0.40,1.50) {Y};
            \node[fill=black!20!green] (v1) at (1.40,-1.65) {X};
            \node[fill=black!20!green] (v2) at (-2.20,-1.65) {A};
            \draw [->] (v1) edge (v0);
            \draw [->] (v2) edge (v0);
        \end{tikzpicture}
        \label{fig:mcar}
    }
        }\hfill 
    \subfloat[MAR]{
    \scalebox{0.76}{
        \begin{tikzpicture}
            \node[color=black!40!gray] (m) at (-2.40,1.50) {M$_Y$};
            \node[fill=black!20!yellow] (v0) at (-0.40,1.50) {Y};
            \node[fill=black!20!green] (v1) at (1.40,-1.65) {X};
            \node[fill=black!20!green] (v2) at (-2.20,-1.65) {A};
            \draw [->] (v2) edge (m);
            \draw [->] (v1) edge (v0);
            \draw [->] (v2) edge (v0);
        \end{tikzpicture}
    }
        \label{fig:mar}
        }\hfill 
    \subfloat[MNAR$^a$]{
    \scalebox{0.76}{
        \begin{tikzpicture}
            \node[color=black!40!gray] (m) at (-2.40,1.50) {M$_Y$};
            \node[fill=black!20!yellow] (v0) at (-0.40,1.50) {Y};
            \node[fill=black!20!green] (v1) at (1.40,-1.65) {X};
            \node[fill=black!20!green] (v2) at (-2.20,-1.65) {A};
            \draw [->] (v0) edge (m);
            \draw [->] (v1) edge (v0);
            \draw [->] (v2) edge (v0);
        \end{tikzpicture}
        \label{fig:mnar1}
    }
        }\hfill 
        \subfloat[MNAR$^b$]{
        \scalebox{0.76}{
        \begin{tikzpicture}
            \node[color=black!40!gray] (m) at (-2.40,1.50) {M$_Y$};
            \node[fill=black!20!yellow] (v0) at (-0.40,1.50) {Y};
            \node[fill=black!20!green] (v1) at (1.40,-1.65) {X};
            \node[fill=black!20!green] (v2) at (-2.20,-1.65) {A};
            \node[fill=black!20!white] (v3) at (-2.80,0.0) {U};
            \draw [->] (v3) edge (m);
            \draw [->] (v1) edge (v0);
            \draw [->] (v2) edge (v0);
            \draw [->] (v3) edge (v2);
        \end{tikzpicture}
    }
        \label{fig:mnar2}
        }
    \caption{Example m-DAGs depicting three different types of missingness mechanisms. Fully and partially observed variables are depicted in green and yellow squares, respectively. We assume $Y$ to be the disease indicator, $A$ the type of treatment, and $X$ the patient age. $M_Y$ is the missingness indicator for the outcome $Y$.}
    \label{fig:missingness}
\end{figure}
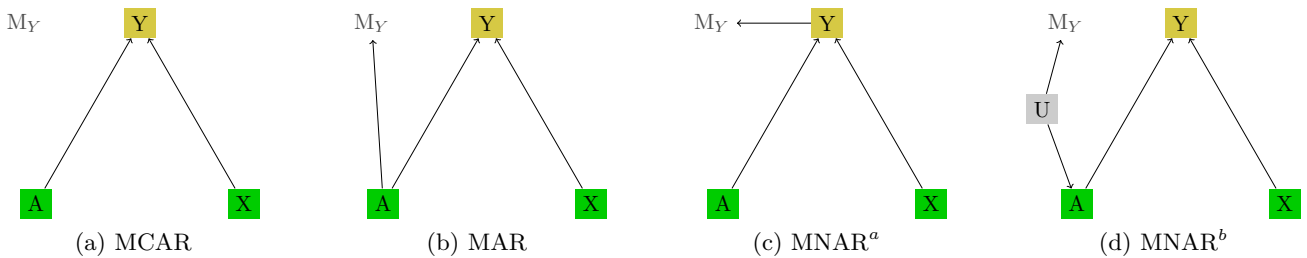

\begin{enumerate}
    \item \textbf{Missing completely at random (MCAR).} If missingness occurs randomly and is assumed not to be caused by any variable in the model, missing or observed, we write $$p(\mathbf{M}|\mathbf{V}_o, \mathbf{V}_m, \mathbf{U}) = p(\mathbf{M}),$$ which means that the conditional distribution of the missingness mechanisms given the variables in the data set and, possibly, also the latent variables, is equal to the marginal distribution of the missingness mechanism. This corresponds to the (unconditional) independence statement $(\mathbf{V}_o, \mathbf{V}_m, \mathbf{U}) \indep \mathbf{M}$.

    In terms of an m-DAG, if the joint distribution $p(\mathbf{V}_o, \mathbf{V}_m, \mathbf{U}, \mathbf{M})$ is faithful with respect to the graph $\mathbf{G}$, this means that there are \emph{no edges between the $\mathbf{M}$ variables and variables in $\mathbf{V}_o$ and $\mathbf{V}_m$}, and parents of missingness indicator variables from $\mathbf{M}$ can only be other variables from $\mathbf{M}$.

    A commonly used example of the MCAR mechanism is an accidental technical error in an electronic medical record, which leads to a loss of data on the disease indicator, compare with Figure \ref{fig:mcar}.

    \item \textbf{Missing at random (MAR).} In this case, the missingness mechanism is assumed to depend on the fully observed variables only, but is not allowed to depend on the missing data values or latent variables. In terms of the conditional distribution of $\mathbf{M}$, this corresponds to $$p(\mathbf{M}|\mathbf{V}_o, \mathbf{V}_m, \mathbf{U}) = p(\mathbf{M}|\mathbf{V}_o).$$

    The conditional independence statement that holds in this case is $(\mathbf{V}_m, \mathbf{U}) \indep \mathbf{M} | \mathbf{V}_o$, which in terms of m-DAGs (if the joint distribution is faithful to $\mathbf{G}$) means that \emph{variables from $\mathbf{M}$ are not allowed to have any parents from the sets $\mathbf{V}_m$ or $\mathbf{U}$, but only from $\mathbf{V}_o$ and other $\mathbf{M}$ variables}.

    For example, if a disease indicator is missing for some patients, and missingness for some reason depends on the completely observed treatment only, one says that the missing data underlies the MAR mechanism, as depicted in the example in Figure \ref{fig:mar}.

    \item \textbf{Missing not at random (MNAR).} This category of missingness is most general. Data that cannot be classified as MCAR or MAR fall in the MNAR category. In this case, the conditional distribution of $\mathbf{M}$, $p(\mathbf{M}|\mathbf{V}_o, \mathbf{V}_m, \mathbf{U})$, cannot be simplified. In an m-DAG (again, assuming faithfulness), if at least one $\mathbf{M}$ variable has a parent which is a latent variable $\mathbf{U}$ or is any of the partially observed variables from the set $\mathbf{V}_m$, then the missing data mechanism is MNAR.

    A common example of MNAR is when a variable is causing its own missingness, e.g., if the missingness of the disease status entry depends on the disease status itself, as this is the case for the m-DAG in Figure \ref{fig:mnar1}. For instance, the presence of a disease is likely to increase the chances of disease status being recorded due to the associated medical follow-ups and tests.

    Another practically very relevant example is presented in Figure \ref{fig:mnar2}. It shows a case where missingness is driven by latent variables, which may potentially also be a cause for an observed variable. For example, socio-economic status (SES) may not be recorded in a medical study, but has an impact on both the missingness in the disease indicator due to reduced healthcare visits and the type of treatment an individual receives, since those with higher SES may have access to better healthcare facilities and more advanced treatment options.
\end{enumerate}

Note that both missing data taxonomies are equivalent if i) observations are independent, and ii) missingness indicators are conditionally independent \citep{schomaker2020}. In this case, the conditional distribution from the graph-based MAR definition can be written as a product of conditional distributions for single missingness indicators as well as for single observations, and this then leads to the equivalence of both definitions.

\subsection{Recoverability of Target Quantities}

In general, recoverability is a property of the target quantity/parameter $\theta$ (and the probability distribution $P_{X}$) which states whether this quantity can be estimated consistently from the available data. To decide on recoverability, one needs to know the independence statements that hold in the joint distribution. However, under the Markovness and faithfulness assumptions, all conditional independence statements present in the data can be directly read off from the m-DAG $\mathbf{G}$, and recoverability can therefore be considered as a property of the pair $\{ \theta, \mathbf{G}\}$ \citep{mohan2021}. Note that the term \emph{identifiability} is used when assessing whether a causal query can be expressed as a function of the observational distribution, whereas the \emph{recoverability} concept is used to describe whether a parameter (not necessarily in a causal context) can be expressed as a function of the available data distribution \citep{moreno2018}.

\citet{moreno2018} describe three main conditions required for recoverability of a target parameter $\theta$: consistency and well-defined interventions, positivity, and conditional independence conditions. In this context, consistency is a property of central relevance. It says that the factual treatment value coincides with the counterfactual outcome.

It has to be emphasized that many types of target parameters $\theta$ may be of interest, and different missingness patterns may occur, which makes it impossible to derive a general automatized algorithm allowing a decision about the recoverability of any specific $\theta$. Therefore, the authors work on `canonical' scenarios which are most typical for epidemiological studies. \citet{mohan2013}, \citet{mohan2021} present some theoretical results on recoverability of joint and conditional distributions. In this work, we only briefly provide intuition on how recoverability works, focusing on aspects that are relevant for our own identification strategies. The main goal is to exploit conditional independencies between the variables of interest and the missingness indicators (which can be read off from the d-separation statements that hold in the graphs) in order to be able to condition on the missingness indicator variables (i.e., on $M_i = 0$ for some $i$). This way, observed variable values will be sufficient for consistent estimation of the target quantity. \\

In the following, we focus on the recoverability of causal effects. Indeed, necessary and sufficient conditions exist for recoverability of causal effects \citep{mohan2021}. Under the presence of missing data, a \emph{necessary condition for recoverability} of a causal effect is the identifiability of this effect from the c-DAG of interest. The causal effect is \emph{identifiable} from a graph $\mathbf{G}$ if the interventional distribution can be determined uniquely from the observed-data distribution alone \citep{pearl2009, tian2002}. \citet{algo_shpitser} provide a sound and complete algorithm for conditional causal effect identification (IDC), which, for any causal effect, can be used for determining identifiability and also for generation of an expression for the interventional distribution in the case of an identifiable effect. \citet{causaleffect_package} implemented the IDC algorithm in the \texttt{R}-package \texttt{causaleffect}. Using the IDC algorithm, we get an estimand of the causal query whenever identifiability holds for the causal effect. In turn, a sufficient condition for recoverability is that the (identified) estimand is recoverable from the missingness graph. This, as mentioned before, has to be decided on a case-by-case basis.

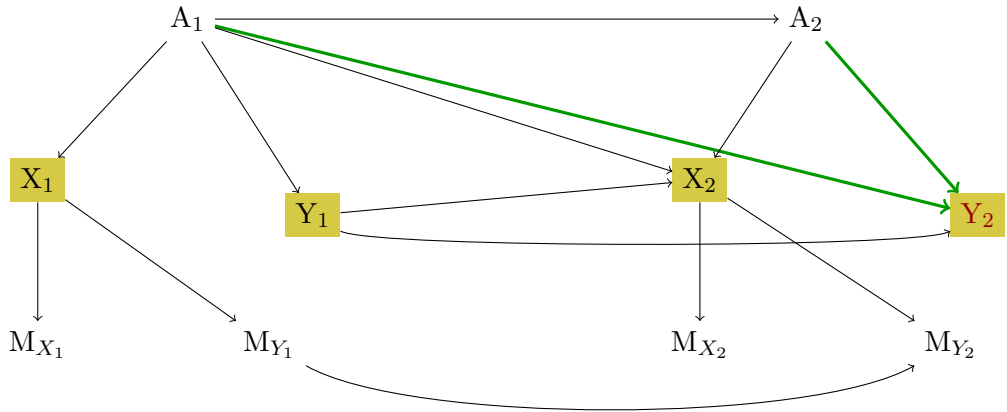
\begin{figure}[ht]
    \begin{center}
        {\noindent
        \scalebox{1}{
            \begin{tikzpicture}[x=2.5in,y=2in]
            \node (v0) at (-1.33,1.1) {A$_1$};
            \node (v1) at (-0.0477,1.1) {A$_2$};
            \node[fill=black!20!yellow] (v2) at (-1.64,0.68) {X$_1$};
            \node[fill=black!20!yellow] (v3) at (-1.07,0.588) {Y$_1$};
            \node[fill=black!20!yellow] (v4) at (-0.268,0.68) {X$_2$};
            \node[color=red!60!black, fill=black!20!yellow] (v5) at (0.308,0.588) {Y$_2$};
            \node (v6) at (-1.64,0.25) {M$_{X_1}$};
            \node (v7) at (-0.268,0.25) {M$_{X_2}$};
            \node (v8) at (0.252,0.25) {M$_{Y_2}$};
            \node (v9) at (-1.16,0.25) {M$_{Y_1}$};
            \draw [->] (v0) edge (v1);
            \draw [->, very thick, color=green!60!black] (v0) edge (v5);
            \draw [->] (v0) edge (v2);
            \draw [->] (v0) edge (v3);
            \draw [->] (v0) edge (v4);
            \draw [->] (v3) edge (v4);
            \draw [->] [bend right, looseness=0.15] (v3) edge (v5);
            \draw [->] (v2) edge (v6);
            \draw [->] (v2) edge (v9);
            \draw [->] [bend right, looseness=0.5] (v9) edge (v8);
            \draw [->] (v4) edge (v7);
            \draw [->] (v4) edge (v8);
            \draw [->] (v1) edge (v4);
            \draw [->, very thick, color=green!60!black] (v1) edge (v5);
            \end{tikzpicture}
            }
        }
    \end{center}
    \caption{An m-DAG depicting the MNAR missingness mechanism in a simple longitudinal study. Health outcomes at two study time points are denoted as $Y_1$ and $Y_2$, sequential treatment variables as $A_1$ and $A_2$, and side effects as $X_1$ and $X_2$.}
    \label{fig:longi_example}
\end{figure}

Next, an example of recoverability procedure of a causal effect for a simple longitudinal study with two measurement points is presented. The example is inspired by \citet[Section 3.5]{mohan2021}. Consider the m-DAG in Figure \ref{fig:longi_example}, which is a model of a simple two-time-point longitudinal study with attrition. The variables $A_1$ and $A_2$ correspond to sequential treatment, $X_1$ and $X_2$ are the side effects, and $Y_1$ and $Y_2$ model some health outcomes. The causal effect of interest is $P(Y_2; do(A_1, A_2))$, which is the impact of the two sequential treatments on the outcome at the second study time point. We can see that the partially observed variables $X_1$ and $X_2$ are causing their own missingness, which means that the missingness mechanism is of the MNAR type. However, even in this case, it can be shown that the causal effect of interest is recoverable using sequential factorization \citep{mohan2021}. First, with the help of the IDC algorithm, an expression for identifying the causal effect from the observable data is provided (for the case as if there had been no missingness, which is indicated in terms of potential outcomes): {\small{$$P(Y_2^{M=0}; do(A_1, A_2)) = \sum_{Y_1} P(Y_2^{M=0}| A_1, A_2, Y_1^{M=0}) \times P(Y_1^{M=0}|A_1).$$}}
Note that the potential outcome notation $(\cdot)^{M=0}$ is explicitly used only for the variables observed incompletely. In the next step, it has to be decided on the recoverability of the two conditional distributions $P(Y_2^{M=0}| A_1, A_2, Y_1^{M=0})$ and $P(Y_1^{M=0}|A_1)$.
{\small{
\begin{flalign*}
    P(Y_2^{M=0}| A_1, A_2, Y_1^{M=0}) &=  P(Y_2^{M=0}| A_1, A_2, Y_1^{M=0}, M_{Y_1} = 0, M_{Y_2} = 0) && \text{$(Y_2 \indep \{M_{Y_1}, M_{Y_2}\} | A_1, A_2, Y_1)$} \\
    &= P(Y_2| A_1, A_2, Y_1, M_{Y_1} = 0, M_{Y_2} = 0) && \text{(by consistency)}
\end{flalign*}
\begin{flalign*}
    P(Y_1^{M=0}|A_1) &=  P(Y_1^{M=0}|A_1, M_{Y_1} = 0) && \text{$(Y_1 \indep M_{Y_1} | A_1)$} \\
    &= P(Y_1|A_1, M_{Y_1} = 0) && \text{(by consistency)}
\end{flalign*}
}}

Formally, it is true that $\{A_1, A_2, Y_1\}$ is the minimal set for which
$$Y_2 \indep \{X_1, X_2\} | A_1, A_2, Y_1.$$ Analogously, $A_1$ is the minimal set for which holds $$Y_1 \indep X_1|A_1.$$
Further, $P(Y_2| A_1, A_2, Y_1)$ satisfies $Y_2 \indep \{M_{Y_1}, M_{Y_2}\} | A_1, A_2, Y_1$, whereas for $P(Y_1|A_1)$ it holds that true $Y_1 \indep M_{Y_1}|A_1$. Then, applying the sequential factorization technique, both factors can be recovered as $P(Y_2| A_1, A_2, Y_1, M_{Y_1} = 0, M_{Y_2} = 0)$ and $P(Y_1|A_1, M_{Y_1} = 0)$, correspondingly.

This shows that the causal effect of interest can be recovered from the available data only, despite the fact that the underlying missingness mechanism is MNAR. \\

After a causal effect of interest has been identified as a function of the observed data, (and after it has been further ensured that and how it can be recovered in the presence of missing data), the causal effect can be estimated using the common techniques, e.g., (parametric) G-computation \citep{robins1986}, inverse probability weighting \citep{hernan2006, hernan2020}, or targeted maximum likelihood estimation (TMLE) \citep{laan2011targeted}.

\subsection{Closed Missingness Mechanism}

Next, an interesting special case of a missingness mechanism, which we refer to as \emph{closed missingness mechanism}, is discussed. We present the recoverability results for joint and conditional distributions under the closed missingness mechanism.

\begin{definition}[Closed missingness] \label{def:closed_miss}
    Consider a c-DAG $\mathbf{G}_c$ with the node set $\mathbf{V} = \{X_1, X_2, ..., X_n\}$, $|\mathbf{V}|=n$. Without loss of generality, assume that $\mathbf{V}_o = \{X_1, ..., X_k\}$ and $\mathbf{V}_m = \{X_{k+1},..., X_n\}$ for some $k \leq n$. The corresponding m-DAG consists of the set $\mathbf{V}$, the missingness indicator variable set $\mathbf{M}=\{M_{X_{k+1}},..., M_{X_n}\}$, and possibly a variable set $\mathbf{Z}$ containing auxiliary variables that are causes of missingness. The missingness mechanism is called closed if there is no path between any $V_i \in \mathbf{V}$ and any $M_{V_i} \in \mathbf{M}$, $i \in \{k+1, ..., n\}$.
\end{definition}

In other words, the missingness mechanism is closed if only the auxiliary variables from $Z$ are causes of missingness. From a practical standpoint, this type of missingness mechanism is common in clinical, epidemiological, and pharmacological studies. The variables of interest in these studies are often of biological and medical nature (e.g., medication/therapy type, medication dose, blood values), whereas the causes of missingness in such variables are typically related to external factors such as technical issues with medical devices and the socioeconomic status of study participants.

\begin{corollary}
    Consider an m-DAG $\mathbf{G}$ depicting a closed missing mechanism, and the general setting as in Definition \ref{def:closed_miss}. The joint distribution $P(\mathbf{V}_o, \mathbf{V}_m)$, and therefore also any marginal or conditional distribution, is recoverable from the available cases.
    
    \begin{proof}
    As there is no path between any $V_i \in \mathbf{V}, i \in \{1,...,n\}$, and any $M_{V_i} \in \mathbf{M}$, $i \in \{k+1, ..., n\}$, $V_i$ and $M_{V_i}$ are d-separated for any $i$ without conditioning on any other variables, and therefore $V_i \indep M_{V_i}$ holds true.
\end{proof}
\end{corollary}

Note that the variables from $\mathbf{Z}$ may also be completely and incompletely observed, or even unobserved or latent, but this plays no role for the variables of interest from the set $\mathbf{V}$. \\

The concept of a `closed missingness mechanism' can lead to a useful practical rule of thumb, especially under MNAR. If a researcher determines that missingness is likely caused by unmeasured factors, they may initially conclude that the data are MNAR, making both imputation and available case analyses invalid. However, by sketching a c-DAG and m-DAG (incorporating missingness indicators), if no arrows are found from the c-DAG to the m-DAG, they can conclude that a available or complete case analysis is valid, despite MNAR. Intuitively, if the causes of missing values are unrelated to the variables necessary for identification, then relying solely on complete cases might be preferable to imputation.

This approach underscores a crucial practical insight: even in the presence of MNAR, if the missingness mechanism is closed, it does not impact the validity of complete case analysis. Thus, researchers can confidently use complete case data when the missingness indicators are isolated from the primary variables of interest.

\section{Data Analysis}\label{sec4}

\subsection{Study setting, measurements and estimand} \label{subsec:data}
As described in Section \ref{sec2}, our data comes from the study of \citet{bienczak2016} and includes 125 children with HIV who were enrolled in the CHAPAS-3 trial and treated with an efavirenz-based regimen. We consider the trial's follow-up visits at 6, 36, 48, 60, and 84 weeks.

Our scientific goal is to estimate causal concentration response-curves. That is, we are interested in the counterfactual probability of a viral load (VL) $>100$ copies/ml (which is considered to be a viral failure) at 36 and 84 weeks if children had efavirenz concentrations (12h after dose) of $x$ mg/L at each follow-up visit, where $x$ ranges from 0 to 10 mg/L. Missing data in the outcome (VL), efavirenz exposure (EFV) and time-dependent confounders (weight, adherence) complicates this exercise.

Measured baseline variables include sex, age, the nucleoside reverse transcriptase inhib\-itors drug (NRTI) component of the treatment regimen and weight. Moreover, we include knowledge on the metabolism status (``genotype'', slow, intermediate, extensive) related to the single nucleotide polymorphisms in the CYP2B6 gene, which is relevant for metabolizing evafirenz and directly affects its concentration in the body. Additionally available data include knowledge on caregiver's beliefs in the necessity of medicine (BMQ, \citealp{Abongomera:2017}), as well as socio-economic indicators (SES). Measured follow-up variables are weight, adherence (measured through memory caps, MEMS) and dose.

Clinical assessments were made at every visit, viral loads were measured at all time points except week 6, efavirenz levels were measured at all assessments other than week 48, and assessment of adherence through MEMS primarily at weeks 36, 48 and 60 as both funding constraints and practical considerations did not allow its consecutive implementation.

\subsection{Development of the causal model}
Figure \ref{fig:full_m-DAG} contains our proposed causal model. The causal graph contains 1) variables important for identifying the effect of interest (i.e. the impact of EFV on viral failure, c-DAG, bottom). We summarized 2) clinician’s knowledge on why data are possibly missing in a m-DAG using binary missingness indicators (top, pink shading) for relevant variables with missing data (EFV, elevated viral load (VL), adherence (MEMS), weight). The corresponding non-parametric structural equation models are given in Appendix \ref{app:scm}.

\subsubsection{c-DAG}
The c-DAG summarizes our knowledge and assumptions, described in more detail in \citet{Schomaker:2023b}. Briefly, the primary cause of viral failure is subtherapeutic efavirenz concentration (EFV$_t$ $\rightarrow$ VL$_t$). The concentration itself depends on the administered dose (Dose$_t$ $\rightarrow$ EFV$_t$), adherence to treatment (MEMS$_t$ $\rightarrow$ EFV$_t$) and how fast the drug is cleared, determined - amongst other - by the 516G and 983T polymorphisms in the CYP2B6 gene (Genotype $\rightarrow$ EFV$_t$). Both weight and MEMS are assumed to be time-dependent confounders. Co-morbidities (CoMo), which we defined to include tuberculosis, pneumonia and other AIDS-defining events are reflected in the DAG, although they are expected to be less frequent in our population, as children with active infections, those being treated for tuberculosis and with severe laboratory abnormalities at screening, were not enrolled into the study, and most children were in relatively good health (e.g. median CD4 cell percentage is 19\%).

\subsubsection{m-DAG}
\textbf{Main m-DAG} ($\mathbf{G}_{main}$): The development of the missingenss causal model is a novel contribution of our paper. Reasons for missing data were obtained from the trial team and the paediatricians. Those reasons are represented by arrows leading into the missingness indicators and include  i) technical issues with the memory caps (e.g., broken containers) or in obtaining or analysing a blood sample (TI, unmeasured); ii) missed visits (MV), which are likely related to socio-economic status of caregivers (SES, measured), beliefs and attitudes towards medicine (BMQ, measured) and other behavioural factors (BHV, unmeasured). As almost all children depend on their caregiver to arrive at their appointment, we assume that the children's age does not affect the probability of a missed visit.

\textbf{Sensitivity of the main causal missingness model} ($\mathbf{G}_{alt1}$): Although the clinicians did not report any other possible reasons for missingness, we added another speculative reason for missed visits to explore the implications of deviations from the assumed causal model: health status of the child, captured indirectly through elevated viral load. This potential reason is represented by the blue dashed arrows in the m-DAG ($\mathbf{G}_{alt1}$). In the discussion, we also mention another m-DAG ($\mathbf{G}_{alt2}$), where $\mathbf{G}_{main}$ is extended with additional arrows SES $\rightarrow$ MEMS$_t$, $t \in \{6, 36, 48, 60, 84\}$.

\begin{figure}[htbp]
\begin{center}
{\noindent
    \scalebox{0.65}{
    \begin{tikzpicture}[x=7.55in,y=3.5in]
    \node (v0) at (0.10,-0.309) {Age};
    \node (v1) at (0.20,-0.540) {CoMo$_0$};
    \node (v2) at (0.50,-0.540) {CoMo$_{36}$};
    \node (v3) at (0.65,-0.540) {CoMo$_{48}$};
    \node (v4) at (0.35,-0.540) {CoMo$_6$};
    \node (v5) at (0.80,-0.540) {CoMo$_{60}$};
    \node (v6) at (0.95,-0.540) {CoMo$_{84}$};
    \node (v8) at (0.55,-0.420) {Dose$_{36}$};
    \node (v9) at (0.70,-0.420) {Dose$_{48}$};
    \node (v10) at (0.40,-0.420) {Dose$_6$};
    \node (v11) at (0.85,-0.420) {Dose$_{60}$};
    \node (v12) at (1.00,-0.420) {Dose$_{84}$};
    \node[color=green!60!black,fill=black!20!yellow] (v14) at (0.60,-0.280) {EFV$_{36}$};
    \draw[green!60!black, semitransparent, very thick] (0.60,-0.280) ellipse[x radius=16pt, y radius=8pt];
    \node[color=green!60!black,fill=black!20!yellow] (v15) at (0.75,-0.280) {EFV$_{48}$};
    \draw[green!60!black, semitransparent, very thick] (0.75,-0.280) ellipse[x radius=16pt, y radius=8pt];
    \node[color=green!60!black,fill=black!20!yellow] (v16) at (0.45,-0.280) {EFV$_6$};
    \draw[green!60!black, semitransparent, very thick] (0.45,-0.280) ellipse[x radius=16pt, y radius=8pt];
    \node[color=green!60!black,fill=black!20!yellow] (v17) at (0.90,-0.280) {EFV$_{60}$};
    \draw[green!60!black, semitransparent, very thick] (0.90,-0.280) ellipse[x radius=16pt, y radius=8pt];
    \node[color=green!60!black,fill=black!20!yellow] (v18) at (1.05,-0.280) {EFV$_{84}$};
    \draw[green!60!black, semitransparent, very thick] (1.05,-0.280) ellipse[x radius=16pt, y radius=8pt];
    \node (v19) at (0.10,-0.396) {Genotype};
    \node[fill=black!20!yellow] (v20) at (0.400,-0.790) {MEMS$_{36}$};
    \node[fill=black!20!yellow] (v21) at (0.550,-0.790) {MEMS$_{48}$};
    \node[fill=black!20!yellow] (v22) at (0.250,-0.790) {MEMS$_6$};
    \node[fill=black!20!yellow] (v23) at (0.700,-0.790) {MEMS$_{60}$};
    \node[fill=black!20!yellow] (v24) at (0.850,-0.790) {MEMS$_{84}$};
    \node (v25) at (0.13,-0.790) {NRTI};
    \node (v26) at (0.10,-0.488) {Sex};
    \node[fill=black!20!yellow] (v27) at (0.30,-0.125) {VL$_0$};
    \node[fill=black!20!yellow] (v28) at (0.60,-0.125) {VL$_{36}$};
    \node[fill=black!20!yellow] (v29) at (0.75,-0.125) {VL$_{48}$};
    \node[fill=black!20!yellow] (v30) at (0.45,-0.125) {VL$_6$};
    \node[fill=black!20!yellow] (v31) at (0.90,-0.125) {VL$_{60}$};
    \node[color=red!60!black, fill=black!20!yellow] (v32) at (1.05,-0.125) {VL$_{84}$};
    \draw[red, semitransparent, very thick] (1.05,-0.125) ellipse[x radius=16pt, y radius=8pt];
    \node[fill=black!20!yellow] (v34) at (0.15,-0.665) {Weight$_0$};
    \node[fill=black!20!yellow] (v35) at (0.45,-0.665) {Weight$_{36}$};
    \node[fill=black!20!yellow] (v36) at (0.60,-0.665) {Weight$_{48}$};
    \node[fill=black!20!yellow] (v37) at (0.30,-0.665) {Weight$_6$};
    \node[fill=black!20!yellow] (v38) at (0.75,-0.665) {Weight$_{60}$};
    \node[fill=black!20!yellow] (v39) at (0.90,-0.665) {Weight$_{84}$};
    \node (v40) at (0.650,0.70) {BMQ};
    \node (v41) at (0.800,0.70) {SES};
    \node (v42) at (0.950,0.70) {\textcolor{black!40!white}{BHV}};
    \node (v43) at (0.35,0.500) {MV$_{0}$};
    \node (v44) at (0.50,0.500) {MV$_{6}$};
    \node (v45) at (0.65,0.500) {MV$_{36}$};
    \node (v46) at (0.80,0.500) {MV$_{48}$};
    \node (v47) at (0.95,0.500) {MV$_{60}$};
    \node (v48) at (1.10,0.500) {MV$_{84}$};
    \node (v50) at (0.35,0.000) {\textcolor{black!40!white}{TI$_{0}$}};
    \node (v51) at (0.50,0.000) {\textcolor{black!40!white}{TI$_{6}$}};
    \node (v52) at (0.65,0.000) {\textcolor{black!40!white}{TI$_{36}$}};
    \node (v53) at (0.80,0.000) {\textcolor{black!40!white}{TI$_{48}$}};
    \node (v54) at (0.95,0.000) {\textcolor{black!40!white}{TI$_{60}$}};
    \node (v55) at (1.10,0.000) {\textcolor{black!40!white}{TI$_{84}$}};
    \node[fill=pink!40!white] (v57) at (0.40,0.400) {M$_{VL_0}$};
    \node[fill=pink!40!white] (v58) at (0.40,0.300) {M$_{Weight_0}$};
    \node[fill=pink!40!white] (v60) at (0.55,0.400) {M$_{VL_6}$};
    \node[fill=pink!40!white] (v61) at (0.55,0.300) {M$_{Weight_6}$};
    \node[fill=pink!40!white] (v62) at (0.55,0.200) {M$_{EFV_6}$};
    \node[fill=pink!40!white] (v63) at (0.55,0.100) {M$_{MEMS_6}$};
    \node[fill=pink!40!white] (v64) at (0.70,0.400) {M$_{VL_{36}}$};
    \node[fill=pink!40!white] (v65) at (0.70,0.300) {M$_{Weight_{36}}$};
    \node[fill=pink!40!white] (v66) at (0.70,0.200) {M$_{EFV_{36}}$};
    \node[fill=pink!40!white] (v67) at (0.70,0.100) {M$_{MEMS_{36}}$};
    \node[fill=pink!40!white] (v68) at (0.85,0.400) {M$_{VL_{48}}$};
    \node[fill=pink!40!white] (v69) at (0.85,0.300) {M$_{Weight_{48}}$};
    \node[fill=pink!40!white] (v70) at (0.85,0.200) {M$_{EFV_{48}}$};
    \node[fill=pink!40!white] (v71) at (0.85,0.100) {M$_{MEMS_{48}}$};
    \node[fill=pink!40!white] (v72) at (1.00,0.400) {M$_{VL_{60}}$};
    \node[fill=pink!40!white] (v73) at (1.00,0.300) {M$_{Weight_{60}}$};
    \node[fill=pink!40!white] (v74) at (1.00,0.200) {M$_{EFV_{60}}$};
    \node[fill=pink!40!white] (v75) at (1.00,0.100) {M$_{MEMS_{60}}$};
    \node[fill=pink!40!white] (v76) at (1.15,0.400) {M$_{VL_{84}}$};
    \node[fill=pink!40!white] (v77) at (1.15,0.300) {M$_{Weight_{84}}$};
    \node[fill=pink!40!white] (v78) at (1.15,0.200) {M$_{EFV_{84}}$};
    \node[fill=pink!40!white] (v79) at (1.15,0.100) {M$_{MEMS_{84}}$};
    \draw [->] (v0) edge (v1);
    \draw [->] (v0) edge (v2);
    \draw [->] (v0) edge (v3);
    \draw [->] (v0) edge (v4);
    \draw [->] (v0) edge (v5);
    \draw [->] (v0) edge (v6);
    \draw [->] (v0) edge (v25);
    \draw [->] (v0) edge (v34);
    \draw [->] (v1) edge (v4);
    \draw [->] (v1) edge (v22);
    \draw [->] (v1) edge (v30);
    \draw [->] (v1) edge (v37);
    \draw [->] (v2) edge (v3);
    \draw [->] (v2) edge (v21);
    \draw [->] (v2) edge (v29);
    \draw [->] (v2) edge (v36);
    \draw [->] (v3) edge (v5);
    \draw [->] (v3) edge (v23);
    \draw [->] (v3) edge (v31);
    \draw [->] (v3) edge (v38);
    \draw [->] (v4) edge (v2);
    \draw [->] (v4) edge (v20);
    \draw [->] (v4) edge (v28);
    \draw [->] (v4) edge (v35);
    \draw [->] (v5) edge (v6);
    \draw [->] (v5) edge (v24);
    \draw [->] (v5) edge (v32);
    \draw [->] (v5) edge (v39);
    \draw [->] (v8) edge (v9);
    \draw [->] (v8) to [bend right, looseness=1.2] (v14);
    \draw [->] (v9) edge (v11);
    \draw [->] (v9) to [bend right, looseness=1.2] (v15);
    \draw [->] (v10) edge (v8);
    \draw [->] (v10) to [bend right, looseness=1.2] (v16);
    \draw [->] (v11) edge (v12);
    \draw [->] (v11) to [bend right, looseness=1.2] (v17);
    \draw [->] (v12) to [bend right, looseness=1.2] (v18);
    \draw [->] (v14) edge (v28);
    \draw [->] (v14) edge (v29);
    \draw [->] (v14) edge (v31);
    \draw [->] (v14) edge (v32);
    \draw [->] (v15) edge (v29);
    \draw [->] (v15) edge (v31);
    \draw [->] (v15) edge (v32);
    \draw [->] (v16) edge (v28);
    \draw [->] (v16) edge (v29);
    \draw [->] (v16) edge (v30);
    \draw [->] (v16) edge (v31);
    \draw [->] (v16) edge (v32);
    \draw [->] (v17) edge (v31);
    \draw [->] (v17) edge (v32);
    \draw [->] (v18) edge (v32);
    \draw [->] (v19) edge (v14);
    \draw [->] (v19) edge (v15);
    \draw [->] (v19) edge (v16);
    \draw [->] (v19) edge (v17);
    \draw [->] (v19) edge (v18);
    \draw [->] (v20) edge (v3);
    \draw [->] (v20) edge (v14);
    \draw [->] (v20) edge (v21);
    \draw [->] (v21) edge (v5);
    \draw [->] (v21) edge (v15);
    \draw [->] (v21) edge (v23);
    \draw [->] (v22) edge (v2);
    \draw [->] (v22) edge (v16);
    \draw [->] (v22) edge (v20);
    \draw [->] (v23) edge (v6);
    \draw [->] (v23) edge (v17);
    \draw [->] (v23) edge (v24);
    \draw [->] (v24) edge (v18);
    \draw [->] (v26) edge (v1);
    \draw [->] (v26) edge (v19);
    \draw [->] (v26) edge (v34);
    \draw [->] (v27) edge (v4);
    \draw [->] (v27) edge (v30);
    \draw [->] (v28) edge (v3);
    \draw [->] (v29) edge (v5);
    \draw [->] (v28) edge (v29);
    \draw [->] (v29) edge (v31);
    \draw [->] (v30) edge (v2);
    \draw [->] (v30) edge (v28);
    \draw [->] (v31) edge (v6);
    \draw [->] (v31) edge (v32);
    \draw [->] (v34) edge (v4);
    \draw [->] (v34) edge (v37);
    \draw [->] (v35) edge (v3);
    \draw [->] (v35)  to [bend right, looseness=1.2](v8);
    \draw [->] (v35) edge (v36);
    \draw [->] (v36) edge (v5);
    \draw [->] (v36) to [bend right, looseness=1.2] (v9);
    \draw [->] (v36) edge (v38);
    \draw [->] (v37) edge (v2);
    \draw [->] (v37) to [bend right, looseness=1.2] (v10);
    \draw [->] (v37) edge (v35);
    \draw [->] (v38) edge (v6);
    \draw [->] (v38) to [bend right, looseness=1.2] (v11);
    \draw [->] (v38) edge (v39);
    \draw [->] (v39) to [bend right, looseness=1.2] (v12);
    \draw [->] (v40) edge (v43);
    \draw [->] (v40) edge (v44);
    \draw [->] (v40) edge (v45);
    \draw [->] (v40) edge (v46);
    \draw [->] (v40) edge (v47);
    \draw [->] (v40) edge (v48);
    \draw [->] (v41) edge (v43);
    \draw [->] (v41) edge (v44);
    \draw [->] (v41) edge (v45);
    \draw [->] (v41) edge (v46);
    \draw [->] (v41) edge (v47);
    \draw [->] (v41) edge (v48);
    \draw [->] (v42) edge (v43);
    \draw [->] (v42) edge (v44);
    \draw [->] (v42) edge (v45);
    \draw [->] (v42) edge (v46);
    \draw [->] (v42) edge (v47);
    \draw [->] (v42) edge (v48);
    \draw [->] (v43) edge (v57);
    \draw [->] (v43) edge (v58);
    \draw [->] (v44) edge (v60);
    \draw [->] (v44) edge (v61);
    \draw [->] (v44) edge (v62);
    \draw [->] (v45) edge (v64);
    \draw [->] (v45) edge (v65);
    \draw [->] (v45) edge (v66);
    \draw [->] (v46) edge (v68);
    \draw [->] (v46) edge (v69);
    \draw [->] (v46) edge (v70);
    \draw [->] (v47) edge (v72);
    \draw [->] (v47) edge (v73);
    \draw [->] (v47) edge (v74);
    \draw [->] (v48) edge (v76);
    \draw [->] (v48) edge (v77);
    \draw [->] (v48) edge (v78);
    \draw [->] (v50) edge (v57);
    \draw [->] (v51) edge (v60);
    \draw [->] (v51) edge (v62);
    \draw [->] (v51) edge (v63);
    \draw [->] (v52) edge (v64);
    \draw [->] (v52) edge (v66);
    \draw [->] (v52) edge (v67);
    \draw [->] (v53) edge (v68);
    \draw [->] (v53) edge (v70);
    \draw [->] (v53) edge (v71);
    \draw [->] (v54) edge (v72);
    \draw [->] (v54) edge (v74);
    \draw [->] (v54) edge (v75);
    \draw [->] (v55) edge (v76);
    \draw [->] (v55) edge (v78);
    \draw [->] (v55) edge (v79);
    \draw [->, very thick, color=blue, dashed] (v27) to [bend left, looseness=1.2] (v43);
    \draw [->, very thick, color=blue, dashed] (v28) to [bend left, looseness=1.2] (v45);
    \draw [->, very thick, color=blue, dashed] (v29) to [bend left, looseness=1.2] (v46);
    \draw [->, very thick, color=blue, dashed] (v30) to [bend left, looseness=1.2] (v44);
    \draw [->, very thick, color=blue, dashed] (v31) to [bend left, looseness=1.2] (v47);
    \draw [->, very thick, color=blue, dashed] (v32) to [bend left, looseness=1.2] (v48);
\node[color=green!60!black] (intervention) at (1.20,-0.45) {Intervention};
    \draw[green!60!black, semitransparent, very thick] (1.20,-0.45) ellipse[x radius=26pt, y radius=13pt];
\node[color=red!60!black] (outcome) at (1.20,-0.3) {Outcome};
    \draw[red, semitransparent, very thick] (1.20,-0.3) ellipse[x radius=26pt, y radius=13pt];
\node[fill=black!20!yellow] (missing) at (1.20,-0.6) {Missing values};
\node[fill=pink!40!white] (miss_indic) at (1.20,-0.7) {Missing. indicat.};
\draw [black!60!white, dashdotted, very thick] (0.08, -0.065) -- (1.22, -0.065);
\draw [decorate,decoration={brace,mirror,amplitude=10pt},thick, black!60!white] (0.05, -0.09) -- (0.05, -0.81) node [black,midway,xshift=-1.2cm,text=black!60!white,scale=1.25] {c-DAG};
    \end{tikzpicture}
    }
}
\end{center}
\caption{m-DAG. Blue dashed arrows distinguish between $\mathbf{G}_{main}$ (without blue arrows) and $\mathbf{G}_{alt1}$ (with blue arrows). The subgraph below the gray dashed line represents the corresponding c-DAG. Unmeasured variables are highlighted in gray. \rule{\textwidth}{0.4pt} \small{Variable names: Sex - biological sex, Weight - body weight, Age - age (in years), VL - elevated viral load, Dose - efavirenz dose administered, EFV - efavirenz concentration (12h after dose), NRTI - nucleoside reverse transcriptase inhibitors drug, Genotype - metabolism status, MEMS - adherence to treatment (measured through memory caps), CoMo - co-morbidities, MV - missed hospital visit, TI - technical issues (e.g., with blood samples or memory caps), BMQ - beliefs and attitudes towards medicine, SES - socio-economic status of caregiver, BHV - behavioural factors.}}
\label{fig:full_m-DAG}
\end{figure}
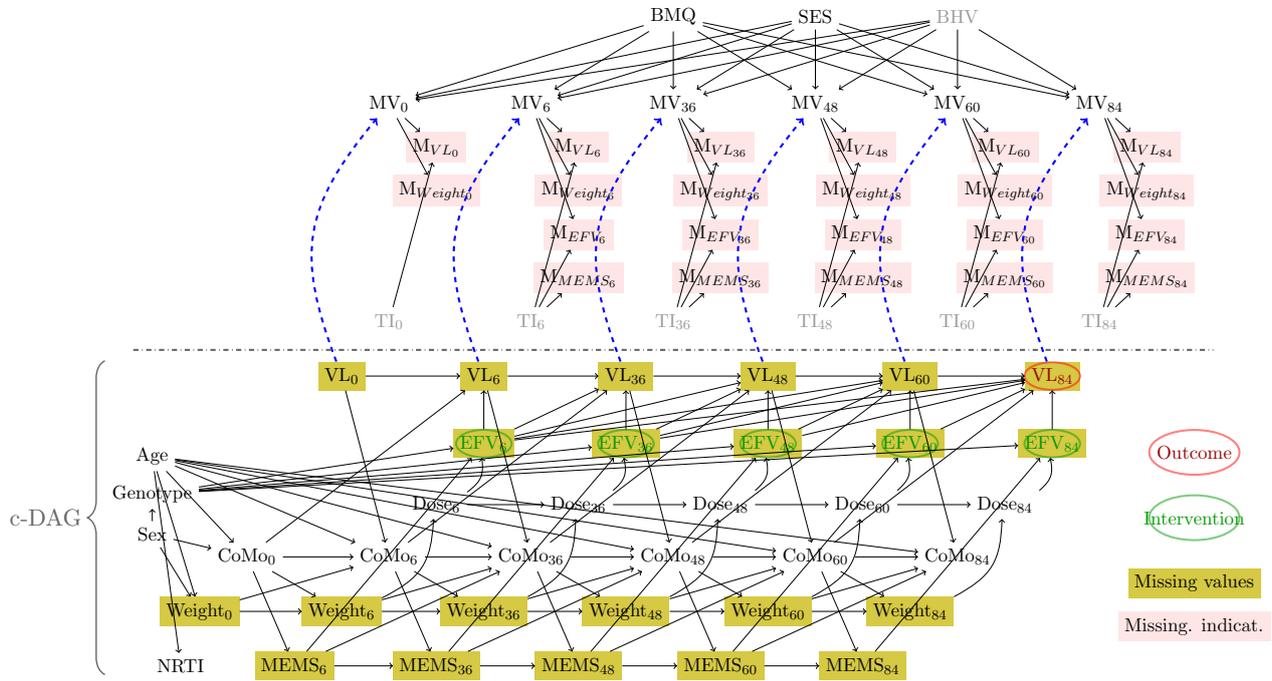

Following the introduced notations, and under the assumption of $\mathbf{G}_{main}$ or $\mathbf{G}_{alt1}$, the sets of completely and partially observed variables, auxiliary variables and missingness indicators are defined as follows:

\begin{itemize}
    \item $\mathbf{V}_o = \{ Age, Genotype, Sex, NRTI, CoMo_t, Dose_t\}$
    \item $\mathbf{V}_m = \{ MEMS_t, Weight_t, EFV_t, VL_t\}$
    \item $\mathbf{M} = \{ M_{VL_t}, M_{Weight_t}, M_{EFV_t}, M_{MEMS_t}\}$
    \item $\mathbf{Z} = \{ TI_t, MV_t, BMQ, SES, BHV\}$.
\end{itemize}
Note that under $\mathbf{G}_{alt2}$ ($\mathbf{G}_{main}$ with additional arrows SES $\rightarrow$ MEMS$_t$), $BHV$ becomes a part of the c-DAG.

\subsection{Identifiability and Recoverability of Causal Effects} \label{subsec:identif}

To make a decision on the recoverability of a causal query of interest, we first need to find out whether the query is identifiable or not. Provided the identifiability holds true, the main idea of recoverability is to transform the partially observed variables from the identified expression into observables with the help of d-separation statements resulting from the m-DAG. \\

In this work, we aim to estimate the causal impact of the efavirenz drug concentration in the plasma of children with HIV, measured and controlled over a specific time period, on viral failure. \\

Particularly, we focus on two causal effects,
{\small{
\begin{flalign} \label{eq:theta_36}
    \theta_{36} & \coloneqq P(VL_{36}=1; \ do(EFV_6 := a, EFV_{36} := a))
\end{flalign}
}}
and
{\small{
\begin{flalign} \label{eq:theta_84}
    \theta_{84} \coloneqq P(VL_{84}=1; do(EFV_6 \coloneqq a, EFV_{36} \coloneqq a, EFV_{48} \coloneqq a, EFV_{60} \coloneqq a, EFV_{84} \coloneqq a)),
\end{flalign}
}}
corresponding to the probability of viral failure after 36 and 84 weeks under a fixed intervention on the plasma concentration of efavirenz drug at each previous up to current study time point. Note that for assessment of recoverability of the causal effects, only variables from the previous up to current study time point are considered, and we therefore focus on the corresponding subgraph of the `full' m-DAG in Figure \ref{fig:full_m-DAG} containing the variables up to and including the 36-th study week when deciding about identifiability and recoverability of $\theta_{36}$. \\

We first consider the m-DAG $\mathbf{G}_{main}$ from Figure \ref{fig:full_m-DAG}, ignoring the dashed blue lines from $VL_t$ to $MV_t$, $t \in \{0, 6, 36, 48, 60, 84\}$. Note that according to the taxonomy of the missingness mechanisms proposed by \citet{mohan2013}, the missingness is of MNAR type because $TI_t$, $t \in \{0, 6, 36, 48, 60, 84\}$, are unobserved variables directly causing missingness. Under Rubin's definition the data would also be MNAR because units may exhibit missing values due to unmeasured behavioral factors (BHV).   \\

In the first step, we have to work on identifiability of the causal effect of interest. We carry this out through two approaches. Our first proposal is the application of the IDC algorithm. Applying rules of \emph{do-calculus} therefore leads to the causal effect identifiable in terms of the observed data distribution as if there were no missingness in any variable relevant for identification:
{\small{
\begin{flalign} \label{eq:ident_theta36_IDC}
\begin{split}
\theta_{36} = \sum_{\substack{Age,Sex,CoMo_0,Weight_0, \\ VL_0,CoMo_6,VL_6}}
& P(VL_{36}=1|Age,Sex,CoMo_0,Weight_0,Genotype,MEMS_6, Weight_6, Dose_6,   \\
& EFV_6 = a, VL_0, CoMo_6, VL_6, MEMS_{36}, Weight_{36}, Dose_{36}, EFV_{36} = a) \\
& P(VL_6|Age,Sex,CoMo_0,Weight_0,Genotype,MEMS_6, Weight_6,\\
& Dose_6,EFV_6 = a,VL_0) \\
& P(CoMo_6|Age,Sex,CoMo_0,Weight_0,Genotype,VL_0) \\
& P(Weight_0|Age,Sex) P(CoMo_0|Age,Sex)\\
&  P(Sex)P(Age)P(VL_0) .
\end{split}
\end{flalign}
}}
The corresponding identifiability result for $\theta_{84}$ based on IDC algorithm is in Appendix \ref{app:identifiability84}. \\

There may be other identifiability results, based on different factorizations. For example, in our second proposal, we apply Pearl's (generalized) back door-criterion to determine a sufficient adjustment set and then simply apply Robins' $g$-formula factorization to it. Such a factorization may involve more conditional distributions that have to be estimated compared to the first approach. For the above estimand, $\theta_{36}$, a valid factorization would be as follows (see, e.g., \citet[Chapters~19~and~21]{HernanRobins2020}):
{\small{
\begin{flalign} \label{eq:ident_theta36_gformula}
\begin{split}
\theta_{36} = \sum_{\substack{Age,Genotype,Sex,NRTI,Weight_0, VL_0 \\ Weight_6,MEMS_6,Weight_{36},MEMS_{36}}}
& P(VL_{36}=1|EFV_6 = a, EFV_{36} = a, Age,Genotype,Sex,NRTI, \\
& Weight_0,Weight_6,MEMS_6,Weight_{36},MEMS_{36}, VL_0, VL_6) \\
& P(Weight_{36}|EFV_6 = a, Age,Genotype,Sex,NRTI, \\
& VL_0,VL_6,Weight_0,Weight_6,MEMS_6) \\
& P(MEMS_{36}|EFV_6 = a, Age,Genotype,Sex,NRTI, \\
& VL_0,VL_6,Weight_0,Weight_6,MEMS_6) \\
& P(VL_{6}|EFV_6 = a, Age,Genotype,Sex,NRTI,VL_0,Weight_0) \\
& P(Weight_{6},MEMS_{6}, Age,Genotype,Sex,NRTI,VL_0,Weight_0)
\end{split}
\end{flalign}
}}

This is because neither the dose nor co-morbidities variables are necessarily required to block the relevant back-door paths from $EFV_t$ to $VL_{t^{\ast}}$, $t^{\ast}\geq t$, i.e. those back-door paths that do not go through any future concentrations.

Note that the identified expressions for $\theta_{36}$ and $\theta_{84}$ are identical under $\mathbf{G}_{main}$ and $\mathbf{G}_{alt1}$ because the c-DAG is identical in both situations. Under assumption of $\mathbf{G}_{alt2}$, the c-DAG changes by addition of the variable $BHV$, resulting in a slight change of the identified expression (see Appendix \ref{subsec:ident_Galt2}). \\

In the second step, we need to decide on the recoverability of the identified expression in terms of the observed data. Note that the missingness mechanism depicted in $\mathbf{G}_{main}$ refers to the \emph{closed missingness mechanism} introduced in the previous section. In this case, there is no path between any variable from the c-DAG containing the variables of interest and the set of missingness indicators $M$ and their causes. Therefore, both causal effects are recoverable, and an available case analysis is admissible in this situation. In order to provide a better intuition about how to decide on the recoverability of a causal effect, we present the recoverability result for $\theta_{36}$ in Appendix \ref{subsec:recover_theta36_Gmain}, implicitly referring to a situation for which no recoverability result exists, and it has to be decided on recoverability manually. \\

If the m-DAG $\mathbf{G}_{main}$ reflects relationships in the data correctly, then an available case analysis is sufficient for estimating the causal effects of interest consistently. However, it is also necessary to assess the plausibility of the graph structure assumption. A suggestion would be to perform a sensitivity analysis to investigate the recoverability of the causal effects $\theta_{36}$ and $\theta_{84}$ under the assumption of an alternative m-DAG which is eligible in terms of content. Therefore, we consider another m-DAG, $\mathbf{G}_{alt1}$, which is depicted in Figure \ref{fig:full_m-DAG} when including the dashed blue lines from $VL_t$ to $MV_t$, $t \in \{0, 6, 36, 48, 60, 84\}$. The arrows from $VL_t$ (binary variable \emph{elevated viral load}) to $MV_t$ (binary variable \emph{missed visit}) for the respective measurement time point reflect the speculation that children with viral failure may miss their appointments due to their poor health condition.

We first investigate the recoverability of $\theta_{36}$ under the assumption of $\mathbf{G}_{alt1}$ being the true underlying m-DAG. Note that if at least one of the conditional distributions in Equation \ref{eq:ident_theta36_IDC} is non-recoverable, we conclude the non-recoverability of the causal query of interest.

We first focus on the second last conditional distribution in Equation \ref{eq:ident_theta36_IDC} which involves the partially observed data, $P(VL_{6}|EFV_6 = a, Age,Genotype,Sex,NRTI,VL_0,Weight_0)$, and the goal is to condition on $M_{Weight_0}=0$, $M_{VL_0}=0$, $M_{EFV_6}=0$ and $M_{VL_6}=0$. From the m-DAG (assuming faithfulness), we know that $VL_6 \indep (M_{Weight_0}, M_{VL_0}, M_{EFV_6}, M_{VL_6})|(MV_0, MV_6)$ (and we always need to condition on (at least) $MV_0$ and $MV_6$ to achieve (conditional) independence), which in turn leads to
{\small{
\begin{flalign*}
    &P(VL_{6}^{M=0}|EFV_6^{M=0} = a, Age,Genotype,Sex,NRTI,VL_0^{M=0},Weight_0^{M=0}) \\
    & = \sum_{MV_0, MV_6} P(VL_{6}^{M=0}|EFV_6^{M=0} = a, Age,Genotype,Sex,NRTI,VL_0^{M=0},Weight_0^{M=0}) \times \\
    & P(MV_0, MV_6|EFV_6^{M=0} = a, Age,Genotype,Sex,NRTI,VL_0^{M=0},Weight_0^{M=0}) \\
    & = \sum_{MV_0, MV_6} P(VL_{6}^{M=0}|EFV_6^{M=0} = a, Age,Genotype,Sex,NRTI,VL_0^{M=0},Weight_0^{M=0},  \\
    & M_{Weight_0}=0, M_{VL_0}=0, M_{EFV_6}=0, M_{VL_6}=0) \times \\
    & P(MV_0, MV_6|EFV_6 = a^{M=0}, Age,Genotype,Sex,NRTI,VL_0^{M=0},Weight_0^{M=0}) \\
    &= \sum_{MV_0, MV_6} P(VL_{6}|EFV_6 = a, Age,Genotype,Sex,NRTI,VL_0,Weight_0, \\
    & M_{Weight_0}=0, M_{VL_0}=0, M_{EFV_6}=0, M_{VL_6}=0) \times \\
    & P(MV_0, MV_6|EFV_6 = a^{M=0}, Age,Genotype,Sex,NRTI,VL_0^{M=0},Weight_0^{M=0}).
\end{flalign*}
}}
The second factor in the expression above, {\small{$$P(MV_0, MV_6|EFV_6^{M=0} = a, Age,Genotype,Sex,NRTI,VL_0^{M=0},Weight_0^{M=0}),$$}} cannot be further decomposed following the adjustment formula, as among others we would need $MV_6$ to be conditionally independent of $M_{EFV_6}$ (possibly given some other variables), but there is no such subset because $MV_6$ is a parent of $M_{EFV_6}$. \\

Because other available recoverability techniques, like sequential factorization or R factorization \citep{mohan2021}, are not applicable in our case, we conclude that the conditional distribution cannot be expressed in terms of the observed data distribution only; this results in non-recoverability of $\theta_{36}$. Based on the same arguments, non-recoverability of $\theta_{84}$ also follows.


\subsection{Analysis}

The analysis is conducted using the m-DAG $\mathbf{G}_{main}$ from Figure \ref{fig:full_m-DAG} and the study data described in Section \ref{subsec:data}. The plug-in g-formula estimation (see, e.g., Hernan and Robins \cite[Chapter 13]{HernanRobins2020}) of our estimands \ref{eq:theta_36} and \ref{eq:theta_84} was based on equations \ref{eq:ident_theta36_gformula} and \ref{eq:ident_theta84_gformula}, respectively. Given the recoverability under $\mathbf{G}_{main}$, we conduct the analysis based on available cases. For modeling the respective conditional distributions, between $69$ and $85$ complete observations are available.

Note that in our setting, relevant co-morbidities in the DAG refer to AIDS-defining events, including tuberculosis, persistent diarrhea, malnutrition, and severe wasting, among others. Most of those events were not measured regularly in our study and thus were not included in the analysis. However, those co-morbidities are not expected to be very frequent in the study population. This is because children with active infections, those being treated for tuberculosis and with severe laboratory abnormalities at screening, were not enrolled into the study, and most children were in relatively good health \citep{mulenga2016}. Also, not all valid identification formulae require information on them, such as those used in our analysis. As adherence could not be measured regularly, as explained in Section \ref{subsec:data}, we constructed a variable indicating any signs of non-adherence, defined as the mean memory caps opening percentage being less than 90\%. 

Although the main analysis was based on available cases due to the identification results, we also implemented a multiple imputation analysis. We consider the results of the multiple imputation analysis invalid because the underlying missingness mechanism is assumed to be MNAR. This is primarily due to the unobserved variables ($TI_t$), which directly cause missingness in variables from the c-DAG.

We multiply imputed five data sets using the Expectation-Maximization Bootstrap (EMB) Algorithm, a joint modeling-based imputation approach implemented in the $R$-package \texttt{Amelia II}. Our setup considered the longitudinal setup through lag-variables and the inclusion of splines of time. Additionally, to improve numerical stability of the EM algorithm, we added a ridge prior (which shrinks the covariances among the variables toward zero without changing the means or variances). This is often recommended when using EMB on small samples  \citep{Honaker:2011}. To address the fact that age, weight and EFV concentrations can not have negative or other illogical values, we specified lower and upper bounds for those variables in the EMB algorithm: $0$ and $35$ for EFV, $0$ and $3$ for $\log$ age and $2$ and $4$ for $\log$ weight. \texttt{Amelia II} implements these bounds by rejection sampling: when drawing the imputations from the respective posterior distributions, all logical constraints need to be satisfied; otherwise, imputations are redrawn until those constraints are met.

\begin{figure}[!ht]
    \centering
    \subfloat[$\theta_{36}$ under $\mathbf{G}_{\text{main}}$]{
        \includegraphics[width=0.48\textwidth,height=0.8\textheight,keepaspectratio]{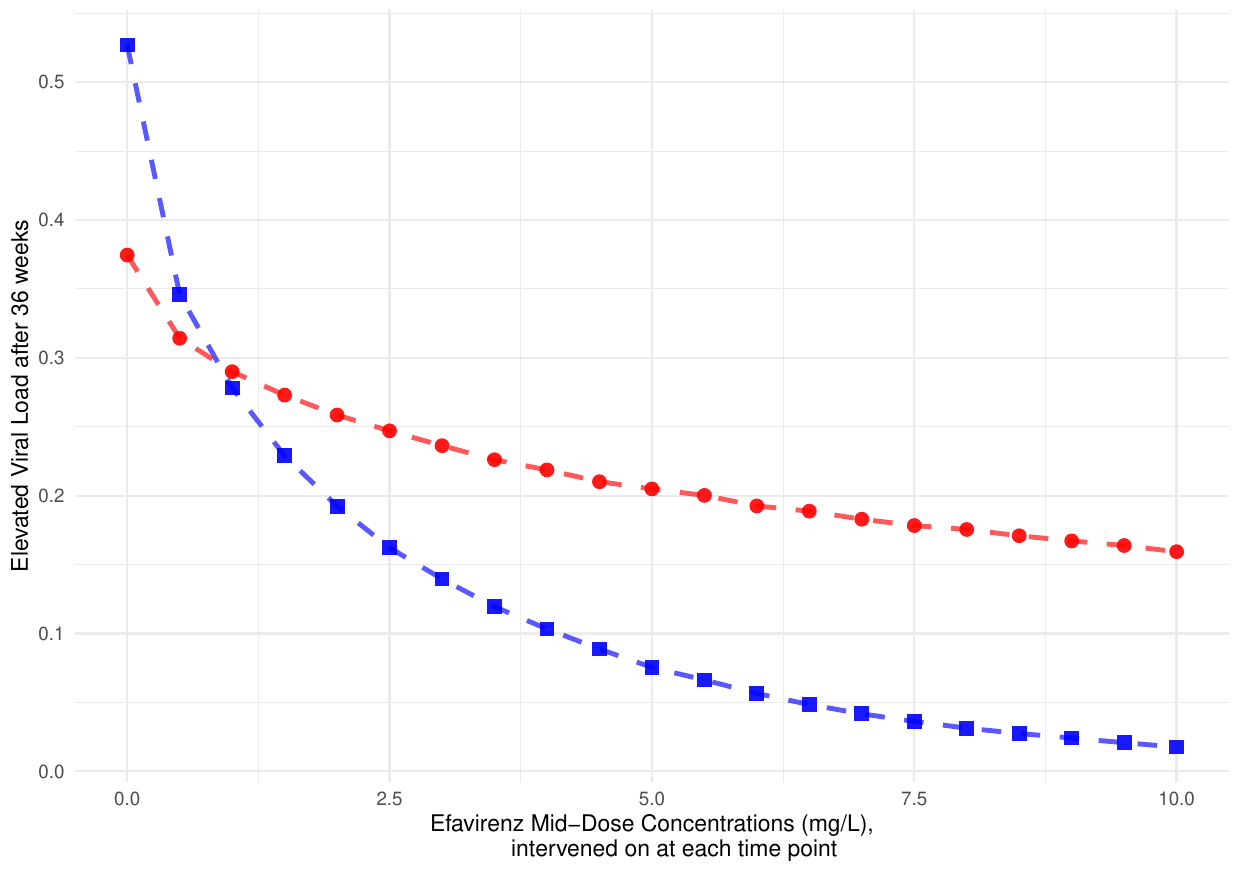}
        \label{fig:chapas_vl36}
    }\hfill
    \subfloat[$\theta_{84}$ under $\mathbf{G}_{\text{main}}$]{
        \includegraphics[width=0.48\textwidth,height=0.8\textheight,keepaspectratio]{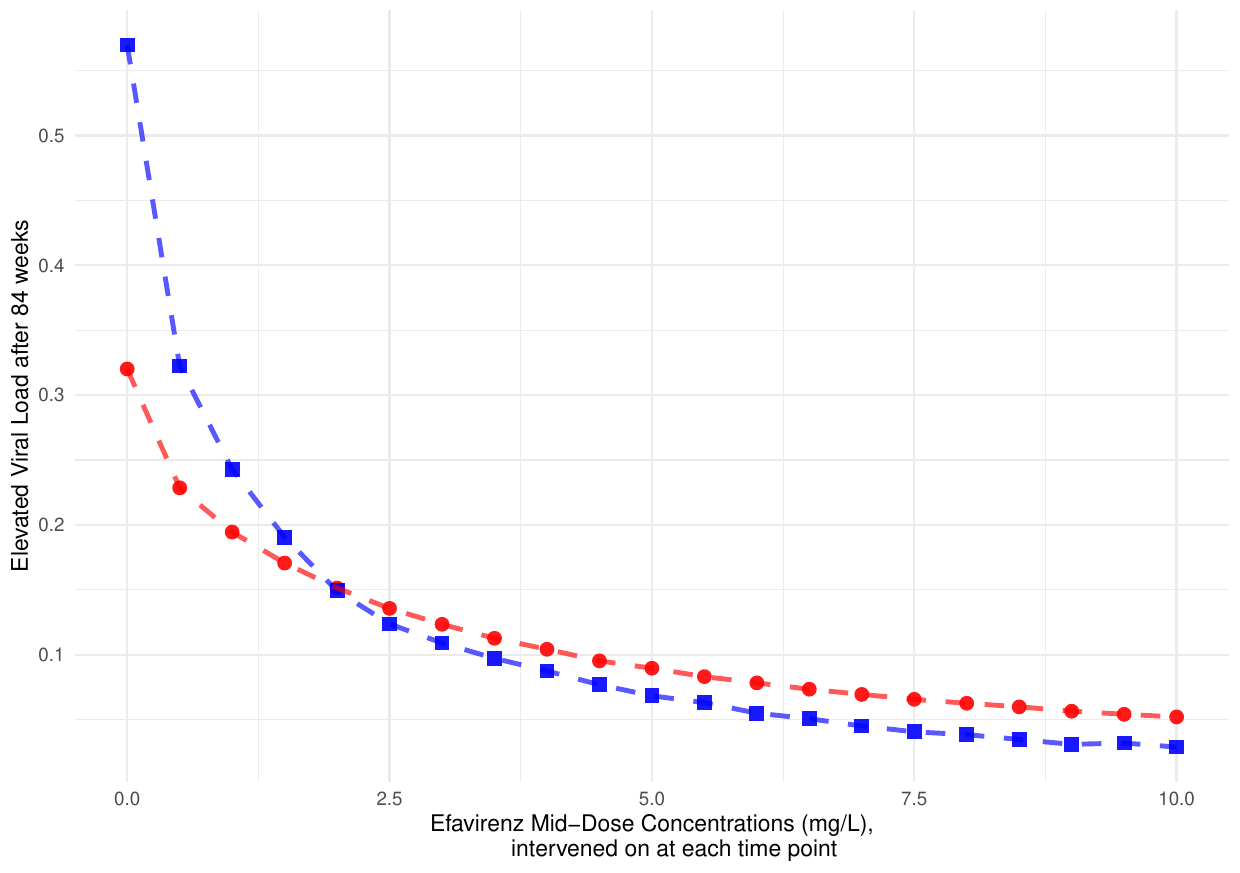}
        \label{fig:chapas_vl84}
    }
    \caption{Estimated CCRCs for $\theta_{36}$ and $\theta_{84}$ under m-DAG $\mathbf{G}_{\text{main}}$. 
    (a) Estimated CCRCs for $\theta_{36}$. (b) Estimated CCRCs for $\theta_{84}$. 
    Causal effects were estimated using available cases (blue squares) and multiple imputation, mean over $20$ imputed data sets (red dots); results represent the mean over $1000$ seeds.}
    \label{fig:chapas}
\end{figure}

The results of our analyses are presented in Figures \ref{fig:chapas_vl36} and \ref{fig:chapas_vl84}.

One can see a higher probability of viral failure with lower EFV concentration values, independent of the approach employed to address the missing data.
Using the suggested available cases approach, the probability of failure is estimated to be $>50$\% at both 36 and 84 weeks if concentrations were $0$mg/L, e.g. if patients did not take any medications. With higher concentrations, failure probabilities decrease to below 5\%, which is expected as EFV is expected to be a potent drug. Interestingly, the CCRCs estimated using multiple imputation are much flatter than under the available case approach, both for $\theta_{36}$ and $\theta_{84}$.

We will now explore the derived theoretical results, along with comparisons between the available case and multiple imputation approaches in the simulation studies below.

\section{Simulation Studies}\label{sec5}

In this section, through simulation studies, we assess the reliability of CCRC estimates for $\theta_{84}$ in the presence of missing data. We compare true CCRC values with estimates derived from: (i) an ideal scenario with no missing data (complete data analysis), (ii) available case analysis, and (iii) multiple imputation.

\subsection{Setup}

The data were simulated using the \texttt{R}-package \texttt{simcausal} \citep{simcausal}, which is a powerful and flexible tool for simulation of longitudinal data structures based on structural causal models. A structural causal model (SCM) uniquely imposes a causal graph, allowing data structures to be generated according to such a causal DAG (in our case, an m-DAG). \\

\emph{Simulation 1:} We simulate data corresponding to both m-DAGs, $\mathbf{G}_{main}$ and $\mathbf{G}_{alt1}$, as defined in Figure \ref{fig:full_m-DAG}. We first simulate binary and normally distributed confounder sets, a continuous (truncated normally distributed) intervention and a binary outcome for all 6 time points and for the sample size of $n = 5.000$. This way, the data aligning with the c-DAG is simulated. Afterwards, we simulate the binary variables causing missingness ($SES_t$ (as proxy for $BMQ_t$, $SES_t$ and $BHV_t$), $MV_t$ and $TI_t$, $t \in \{0, 6, 36, 48, 60, 84\}$. Based on these variables, we simulate the missingness indicators for $VL_t$, $t \in \{0, 6, 36, 48, 60, 84\}$, and $MEMS_t$, $t \in \{6, 36, 48, 60, 84\}$. In this scenario, we assume that only these two variables (for all time points) are subject to missing data. \\

\emph{Simulation 2:} The DGP is the same as for Simulation 1, but we simulate missingness indicators for many more variables: $VL_t$, $Weight_t$, $t \in \{0, 6, 36, 48, 60, 84\}$, and $MEMS_t$, $EFV_t$, $t \in \{6, 36, 48, 60, 84\}$. This scenario mostly coincides with the GPD induced by Figure \ref{fig:full_m-DAG}. This scenario introduces another layer of complexity because the distribution of $EFV_t$, $t \in \{6, 36, 48, 60, 84\}$ is non-symmetric and complex. Using a parametric imputation model with slightly misspecified distributional assumptions may introduce some bias. This setting serves as a benchmark for a realistic scenario where missing data with somewhat complex distributions are imputed using parametric assumptions in conjunction with predictive mean matching (fully conditional imputation procedures, like (M)ICE), or variations thereof (joint modeling, as in \texttt{Amelia II} which we use below).\\

The exact model specifications are given in Appendix \ref{app:simulation_12}. Based on the DGP, we first simulate the interventional data (1000 repetitions) and compute the true CCRC based on it. We intervene on $EFV$ at time points $6, 36, 48, 60,$ and $84$, considering a sequence of interventions from $0$ up to $10$ mg/L with steps of $0.5$ mg/L. Next, the CCRC is estimated using the g-formula factorization  provided in Appendix \ref{app:identifiability84}, once on the complete data and once on the available case data. Compared to the widely used complete case analysis, which relies on samples where all variables in the data set are observed, the available case analysis retains all samples where variables in the query of interest (for the g-formula, the variables required for the estimation of the conditional distribution of interest) are observed. For this reason, an available case analysis is preferred over a complete case analysis due to a more economical usage of available data \citep{mohan2021}. Finally, the CCRC is estimated using multiple imputation (MI) and the g-formula.

\subsection{Results}

\begin{figure}[!ht]
    \centering
    \subfloat[Data simulated under the main DAG $\mathbf{G}_{\text{main}}$]{
        \includegraphics[width=0.48\textwidth,height=0.8\textheight,keepaspectratio]{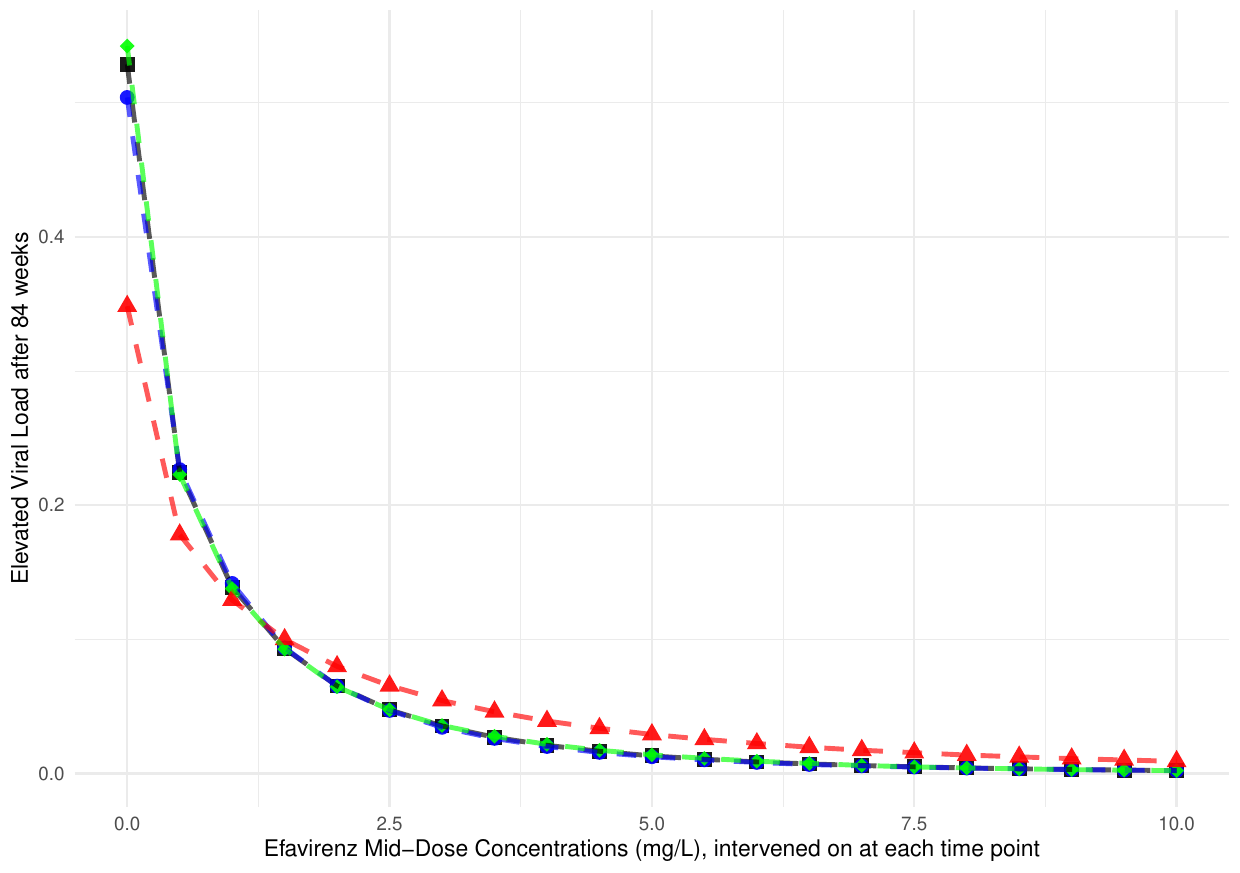}
        \label{fig:CCRC_vl84_main_2miss}
    }\hfill
    \subfloat[Data simulated under the alternative DAG $\mathbf{G}_{\text{alt1}}$]{
        \includegraphics[width=0.48\textwidth,height=0.8\textheight,keepaspectratio]{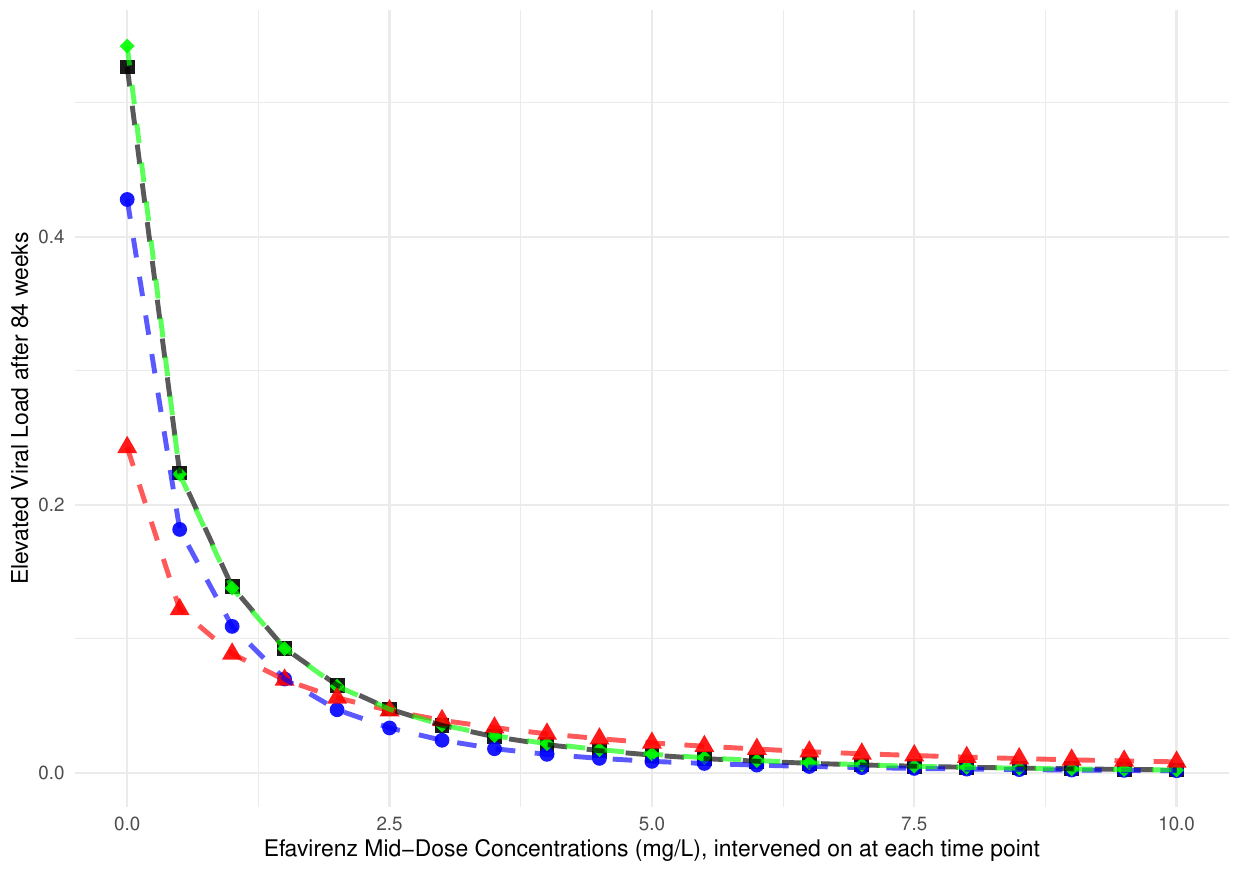}
        \label{fig:CCRC_vl84_altern_2miss}
    }
    \caption{Estimated CCRCs for the probability of viral failure after 84 weeks. (a) Data simulated under the main DAG $\mathbf{G}_{\text{main}}$. (b) Data simulated under the alternative DAG $\mathbf{G}_{\text{alt1}}$. Causal effects were estimated on complete data (black squares), incomplete data using available cases (blue dots), incomplete data using multiple imputation (red triangles) and counterfactual data (green diamonds, true CCRC); results represent the mean over $1000$ seeds.}
    \label{fig:CCRC_vl84}
\end{figure}

Below, we focus on the results from Simulation 1. The results of the Simulation 2are presented in Appendix \ref{app:simulation2}, specifically in Figures \ref{fig:CCRC_vl84_main_4miss} and \ref{fig:CCRC_vl84_altern_4miss}.

The simulation results align with the theoretical findings. We first consider the estimated CCRC curve for elevated viral load after 84 weeks based on data simulated under $\mathbf{G}_{main}$, as presented in Figure \ref{fig:CCRC_vl84_main_2miss}. According to the findings in Section \ref{sec4}, the missingness mechanism is of MNAR type. Despite this, the causal query of interest, $\theta_{84}$, remains recoverable, allowing for consistent estimation of the CCRC via an available case analysis. Conversely, the MI approach is inadmissible due to its underlying assumption of Missing At Random (MAR), which is not met because of the partially observed variables $TI_t$, $t \in {0, 6, 36, 48, 60, 84}$, that affect the probability of missingness in variables from $\mathbf{V}_m$. The estimated counterfactual outcomes, illustrated in Figure \ref{fig:CCRC_vl84_main_2miss}, represent the mean estimated values across $1000$ simulation repetitions. It is evident that the CCRC estimates from available case analysis (dashed blue line) match the true CCRC (dashed green line), whereas those derived from MI differ markedly. To determine whether the discrepancies represent a bias or could also be explained by simulation uncertainty, we calculated Monte Carlo confidence intervals for the differences between the MI estimation results and the true causal effects $\theta_{84}$. The results are reported in Table \ref{tab:mi_main} and show that for each EFV value, the estimated differences are significantly different from zero. This discrepancy underscores the theoretical findings regarding the bias introduced by MI due to the MNAR missingness mechanism. \\

Secondly, consider the results under the alternative DAG $\mathbf{G}_{alt1}$, presented in Figure \ref{fig:CCRC_vl84_altern_2miss}. In this case, the causal query of interest is non-recoverable. The results confirm this and demonstrate that neither the available case analysis (dashed blue line) nor the MI estimates (dashed red line) align with the true CCRC (dashed green line). The respective Monte Carlo confidence intervals, see Tables \ref{tab:ac_alt1} and \ref{tab:mi_alt1} in the appendix, show that these differences cannot be explained by simulation uncertainty alone, which is consistent with the theoretical findings. \\

These findings highlight the critical need to scrutinize the common MAR-type missingness assumption, especially in complex longitudinal data scenarios where multiple variables experience missingness. Our simulation study illustrates how even minor changes in the missingness structure can dramatically affect recoverability and, consequently, the accuracy of estimation results. This emphasizes the importance of careful consideration and justification of missingness assumptions in such analyses.

\section{Conclusions}\label{sec6}

Our analyses demonstrate the applicability of missingness DAGs to complex longitudinal studies and show that, in some cases, available case analyses can be valid under MNAR. However, our application also highlights the massive effort involved, the technical expertise required, and sensitivity of results to the assumed causal model.

Under the assumptions represented in Figure \ref{fig:full_m-DAG}, the data are missing not at random (MNAR) because technical issues ($TI$) with pill containers (frequently) and blood samples (rarely), which we assume to be direct causes of missingness in multiple variables, have not been measured. However, we show that these assumptions are sufficient to estimate the desired impact using the available data and  g-formula representations, despite the MNAR mechanism. However, if reasons for missed visits are caused by the outcome (elevated viral load), as speculated in the alternative DAG, the causal effect cannot be recovered. Interestingly, additional simulations (Appendix \ref{app:simulation2}, Figure \ref{fig:CCRC_vl84_altern2_2miss}) show that recoverability holds true even if behavioural factors directly cause non-adherence. This corresponds to the situation discussed in the second alternative DAG, $G_{alt2}$. However, it is possible that more complex processes exist between socio-economic/behavioural factors and biologic processes, for which identifiability does not hold. Our investigations demonstrate the sensitivity of recoverability results to even minor changes in the missingness structure. We therefore emphasize the need for careful inspection of the assumptions regarding the missing data mechanism, especially in complex longitudinal studies with multiple variables experiencing missingness.

It is also important to highlight that the estimated concentration-response curves are much flatter when using multiple imputation and are actually invalid under the assumptions stated in Figure \ref{fig:full_m-DAG}, as predicted by theory and confirmed by our simulations.

Our analyses show lower probabilities of viral failure with higher concentrationsafter accounting for the missing data. There are, of course, many further complications that may have to be considered for accurate causal inference in our setting. First, one may also account for measurement error, e.g. in viral load and EFV concentrations \citep{Schomaker:2015}, though this may not affect our binary viral failure definition very much. Second, it would be advantageous if measurements were available more frequently and precisely, especially for measuring actual adherence to the prescribed treatment plan, which is difficult in practice.

While advancements in causal modeling and appropriate statistical estimation techniques are impressive, answering complex epidemiological and biological questions remains a challenge.


\subsubsection*{Acknowledgements}
We are grateful for the support of the CHAPAS-3 trial team, their advice regarding the data analysis and making their data available to us. We would like to thank David Burger, Sarah Walker, Di Gibb and Andrzej Bienczak for their help in interpreting the data. We thank Elizabeth Kaudha and Victor Musiime for discussing reasons for missingness with us. We would further like to acknowledge the help of Alexander Szubert in constructing some of the variables.  Michael Schomaker is supported by the German Research Foundations (DFG) Heisenberg Programm (grants 465412241 and 465412441). The CHAPAS-3 trial was funded by the European Developing Countries Clinical Trials Partnership (IP.2007.33011.006), Medical Research Council UK (MC\_UU\_00004/03), Department for International Development UK, and Ministerio de Sanidady Consumo Spain. Cipla Ltd donated first-line antiretrovirals. We thank all the children, carers, and staff from all the centres participating in the CHAPAS-3 trial.

\bibliographystyle{unsrtnat}
{\footnotesize
\bibliography{literature}
}


\clearpage
\appendix

\appendix

\section{Structural equation model} \label{app:scm}

We now present the non-parametric structural equation models (SEMs) corresponding to the m-DAGs in Figure \ref{fig:full_m-DAG}. Note that the follow-up time points $1$ through $5$ coincide with the study weeks $6$, $36$, $48$, $60$ and $84$. The independent (joint) noise term  is denoted as $\mathbf{U}$. \\

The SEM corresponding to $\mathbf{G}_{main}$ in Figure \ref{fig:full_m-DAG} is as follows: \\

{\scriptsize{
\noindent For $t=0$:
\noindent
\begin{eqnarray*}
\text{Genotype} &=& f_{\text{Genotype}}(\text{Sex}, U_{\text{Genotype}}) \\
\text{Weight}_0 &=& f_{\text{Weight}_0}(\text{Sex}, \text{Age}, U_{\text{Weight}_0}) \\
\text{NRTI}_0 &=& f_{\text{NRTI}}(\text{Age}, U_{\text{NRTI}}) \\
\text{MV}_0 &=& f_{\text{{MV}}_0}(\text{BMQ, SES, BHV}, U_{\text{{MV}}_0})
\end{eqnarray*}

\noindent For $t=1$:
\noindent
\begin{eqnarray*}
\text{Dose}_1 &=& f_{\text{Dose}_1}(\text{Weight}_1, U_{\text{Dose}_1})
\end{eqnarray*}

\noindent For $t\geq0$:
\noindent
\begin{eqnarray*}
\text{M}_{Weight_t} &=& f_{\text{M}_{Weight_t}}(\text{MV}_t, U_{\text{M}_{Weight_t}}) \\
\text{M}_{VL_t} &=& f_{\text{M}_{VL_t}}(\text{MV}_t, \text{TI}_t, U_{\text{M}_{VL_t}})
\end{eqnarray*}

\noindent For $t\geq1$:
\noindent
\begin{eqnarray*}
\text{MEMS}_t &=& f_{\text{MEMS}_t}(\text{CoMo}_{t-1}, \text{MEMS}_{t-1}, U_{\text{MEMS}_t}) \quad [\text{assume}  \ \text{MEMS}_0 = 0] \\
\text{Weight}_t &=& f_{\text{Weight}_t}(\text{Weight}_{t-1}, \text{CoMo}_{t-1}, U_{\text{Weight}_t}) \\
\text{CoMo}_t &=& f_{\text{CoMo}_t}(\text{CoMo}_{t-1}, \text{Age}, \text{Weight}_{t-1}, \text{VL}_{t-1}, U_{\text{CoMo}_t}) \\
\text{EFV}_t &=& f_{\text{EFV}_t}(\text{Dose}_t, \text{MEMS}_{t}, \text{Genotype}, U_{\text{EFV}_t}) \\
\text{VL}_t &=& f_{\text{VL}_t}(\text{VL}_{t-1},\text{CoMo}_{t-1}, \text{EFV}_t, U_{\text{VL}_t}) \\
\text{MV}_t &=& f_{\text{MV}_t}(\text{BMQ},\text{SES}, \text{BHV}, U_{\text{MV}_t}) \\
\text{M}_{EFV_t} &=& f_{\text{M}_{EFV_t}}(\text{MV}_t, \text{TI}_t, U_{\text{M}_{EFV_t}}) \\
\text{M}_{MEMS_t} &=& f_{\text{M}_{MEMS_t}}(\text{TI}_t, U_{\text{M}_{MEMS_t}})
\end{eqnarray*}

\noindent For $t\geq2$:
\noindent
\begin{eqnarray*}
\text{Dose}_t &=& f_{\text{Dose}_t}(\text{Dose}_{t-1},\text{Weight}_t, U_{\text{Dose}_t})
\end{eqnarray*}
}}

The SEM for $\mathbf{G}_{alt1}$ (with present blue dashed lines in Figure \ref{fig:full_m-DAG}) is the same as the SEM above, except for the structural equations for $MV_t$, $t \in \{0, 6, 36, 48, 60, 84\}$. These are specified as follows for $\mathbf{G}_{alt1}$: \\

{\scriptsize{
\noindent For $t=0$:
\noindent
\begin{eqnarray*}
\text{MV}_0 &=& f_{\text{MV}_0}(\text{BMQ, SES, BHV}, \text{VL}_0, U_{\text{MV}_0})
\end{eqnarray*}
\noindent For $t\geq1$:
\noindent
\begin{eqnarray*}
\text{MV}_t &=& f_{\text{MV}_t}(\text{BMQ, SES, BHV}, \text{VL}_t, U_{\text{MV}_t})
\end{eqnarray*}
}}
This way, $MV_t$ additionally depends on $VL_t$, $t \in \{0, 6, 36, 48, 60, 84\}$, which corresponds to the dashed blue lines in the DAG. \\

\section{\texorpdfstring{Identifiability results}{Identifiability results}}\label{app:identifiability}
\vspace*{12pt}

\subsection{\texorpdfstring{Identifiability results for $\theta_{84}$}{Identifiability result for theta84}}\label{app:identifiability84}

The identifiability result below is based on application of the IDC algorithm \citep{algo_shpitser}.
{\small{
\begin{flalign} \label{eq:ident_theta84_IDC}
\begin{split}
\theta_{84} =
\sum_{\substack{Age,Sex,CoMo_0,Weight_0, \\
MEMS_6,Weight_6,VL_0,CoMo_6, \\
VL_6,MEMS_{36},Weight_{36},CoMo_{36},\\
MEMS_{48},Weight_{48},VL_{36},CoMo_{48},\\
VL_{48},CoMo_{60},VL_{60}}}
& P(VL_{84}=1|Age,Sex,CoMo_0,Weight_0,Genotype,MEMS_6,Weight_6, \\
& Dose_6,EFV_6 = a,VL_0,CoMo_6,VL_6,MEMS_{36}, \\
& Weight_{36},CoMo_{36},Dose_{36},MEMS_{48},Weight_{48},EFV_{36} = a, \\
& Dose_{48},VL_{36},EFV_{48}=a,CoMo_{48},VL_{48},MEMS_{60},Weight_{60},CoMo_{60}, \\
& Dose_{60},MEMS_{84},Weight_{84},EFV_{60}=a,Dose_{84},VL_{60},EFV_{84}=a) \\
& P(VL_{60}|Age,Sex,CoMo_0,Weight_0,Genotype,MEMS_6,Weight_6, \\
& Dose_6,EFV_6 = a,VL_0,CoMo_6,VL_6,MEMS_{36}, \\
& Weight_{36},CoMo_{36},Dose_{36},MEMS_{48},Weight_{48},EFV_{36} = a,Dose_{48}, \\
& VL_{36},EFV_{48}=a,CoMo_{48},VL_{48},MEMS_{60},Weight_{60},Dose_{60},EFV_{60}=a) \\
& P(CoMo_{60}|Age,Sex,CoMo_0,Weight_0,Genotype,MEMS_6,Weight_6, \\
& Dose_{36},Dose_6,EFV_6 = a,VL_0,CoMo_6,VL_6,MEMS_{36},Weight_{36},CoMo_{36}, \\
& MEMS_{48},Weight_{48}, EFV_{36} = a,Dose_{48},VL_{36},EFV_{48}=a,CoMo_{48},VL_{48})\\
& P(VL_{48}|Age,Sex,CoMo_0,Weight_0,Genotype,MEMS_6,Weight_6,Dose_6, \\
& EFV_6 = a,VL_0,CoMo_6,VL_6,MEMS_{36},Weight_{36},CoMo_{36},Dose_{36}, \\
& MEMS_{48},Weight_{48},EFV_{36} = a,Dose_{48},VL_{36},EFV_{48}=a) \\
& P(CoMo_{48}|Age,Sex,CoMo_0,Weight_0,Genotype,MEMS_6,Weight_6, \\
& Dose_6,EFV_6 = a,VL_0,CoMo_6,VL_6,MEMS_{36}, \\
& Weight_{36},CoMo_{36},Dose_{36},EFV_{36} = a,VL_{36}) \\
& P(Weight_{48}|Age,Sex,CoMo_0,Weight_0,Genotype,MEMS_6,Weight_6, \\
& Dose_6,EFV_6 = a,VL_0,CoMo_6,VL_6,Weight_{36},CoMo_{36}) \\
& P(MEMS_{48}|Age,Sex,CoMo_0,Weight_0,Genotype,MEMS_6,Weight_6, \\
& Dose_6,EFV_6 = a,VL_0,CoMo_6,VL_6,MEMS_{36},CoMo_{36}) \\
& P(CoMo_{36}|Age,Sex,CoMo_0,Weight_0,Genotype,MEMS_6,Weight_6, \\
& Dose_6,EFV_6 = a,VL_0,CoMo_6,VL_6) \\
& P(VL_{36}|Age,Sex,CoMo_0,Weight_0,Genotype,MEMS_6,Weight_6,Dose_6, \\
& EFV_6 = a,VL_0,CoMo_6,VL_6,MEMS_{36},Weight_{36}, Dose_{36}, EFV_{36} = a) \\
& P(Weight_{36}|Age,Sex,CoMo_0,Weight_0,Genotype,Weight_6,\\
& VL_0,CoMo_6)P(MEMS_{36}|Age,Sex,CoMo_0,Weight_0,Genotype,MEMS_6, \\
& VL_0,CoMo_6) P(VL_6|Age,Sex,CoMo_0,Weight_0,Genotype,\\
& MEMS_6,Weight_6, Dose_6,EFV_6 = a,VL_0)\\
& P(CoMo_6|Age,Sex,CoMo_0,Weight_0,Genotype,VL_0) \\
& P(Weight_6|Age,Sex,CoMo_0,Weight_0)P(MEMS_6|CoMo_0) \\
& P(Weight_0|Age,Sex) P(CoMo_0|Age,Sex)P(VL_0)P(Sex)P(Age)
\end{split}
\end{flalign}
}}

An alternative identifiability result can be obtained, for example, by applying the generalized back-door criterion. This results in using an adjustment set given by the confounders weight and adherence (MEMS). Both dose and co-morbidities are not necessarily required to block the relevant back-door paths from $EFV_t$ to $VL_{t^{\ast}}$, $t^{\ast}\geq t$, i.e. the back-door paths that do not pass through any future concentrations \citep{HernanRobins2020}. Applying Robins' parametric g-formula \citep{robins1986, HernanRobins2020} leads to the following valid g-formula factorization:

{\small{
\begin{eqnarray} \label{eq:ident_theta84_gformula}
\theta_{84} &=&
\int_l \left\{ P(VL_{84}=1|\bar{A}_{84}=\bar{a}_{84}, \bar{\mathbf{L}}_{84} = \bar{\mathbf{l}}_{84}) \vphantom{\prod_{s=1}^{q}}\right. \nonumber\\
&& \left. \times \prod_{s=0}^{84} f({\mathbf{L}}_s = {\mathbf{l}}_s | \bar{A}_{s-1}=\bar{a}_{s-1}, \bar{\mathbf{L}}_{s-1} = \bar{\mathbf{l}}_{s-1}) \right\} \, dF_l(l)\,,
\end{eqnarray}
}}
where $F_L(\cdot)$ denotes the CDF with respect to $L$. In the above, $\mathbf{L}_t = \{\text{weight}_t, \text{MEMS}_t\}$, $\mathbf{L}_0 = \{\text{weight}_0, \text{NRTI}, \text{Genotype}, \text{Sex}, \text{Age}\}$ and $A_t = \text{EFV}_t$. Note that the overbar refers to the intervention and confounder histories, i.e. for of a unit $i$ (up to and including time $t$) the histories are$\bar{A}_{t,i}=(A_{0,i},\ldots,A_{t,i})$ and $\bar{L}^s_{t,i}=(L^s_{0,i},\ldots,L^s_{t,i})$, $s=1,2$, $i=1,\ldots,n$, respectively. Outcomes (i.e., VL) prior to $s$ are part of $\bar{\mathbf{L}}_{s}$. The inner product of the $2$ time-varying and ordered confounders ${\mathbf{L}}_s = \{L_s^1,L_s^2\}$  can be decomposed further:
{\small{
\begin{eqnarray*}
\prod_{s=0}^{84} 
f(L_s^2=l_s^2 \mid {L}_s^{1} = {l}_s^{1}, \bar{A}_{s-1}=\bar{a}_{s-1}, {\bar{L}}_{s-1}=\mathbf{\bar{l}}_{s-1}) \times f(L_s^1=l_s^1 \mid  \bar{A}_{s-1}=\bar{a}_{s-1}, \mathbf{\bar{L}}_{s-1}=\mathbf{\bar{l}}_{s-1})
\end{eqnarray*}
}}
Thus, implementing (\ref{eq:ident_theta84_gformula}), requires fitting models for the outcome and two confounder distributions at each time point, given their respective history.

\section{\texorpdfstring{Recoverability results for $\theta_{36}$}{Recoverability results for theta36}}

\subsection{\texorpdfstring{Assuming G\textsubscript{main}}{Assuming Gmain}} \label{subsec:recover_theta36_Gmain}

To assess the recoverability of the identified causal effect $\theta_{36}$ in Equation \ref{eq:ident_theta36_IDC}, we must show that each of the multiplicative factors corresponding to the conditional or marginal distributions can be expressed in terms of the observed data only. Although the missingness mechanism is of the `closed' type, and the recoverability result is directly available for this case, we want to demonstrate how one would examine identifiability if such a result is not available. T do so, we present the result for one of the conditional distributions containing partially observed variables. The results for other multiplicative factors are derived analogously.
{\small{
\begin{flalign*}
&P(CoMo_6^{M=0}|Age,Sex,CoMo_0^{M=0},Weight_0^{M=0},Genotype,VL_0^{M=0}) \\
= &P(CoMo_6^{M=0}|Age,Sex,CoMo_0^{M=0},Weight_0^{M=0},Genotype,VL_0^{M=0}, \mathbf{M}=0) \\
= &P(CoMo_6|Age,Sex,CoMo_0,Weight_0,Genotype,VL_0, \mathbf{M}=0)
\end{flalign*}
}}

The first equality holds due to the fact that all partially observed variables are independent of the corresponding relevant missingness indicators, whereas the second equality is true due to the consistency assumption \citep{moreno2018}.

Note that the recoverability result above is trivial due to the specific missingness mechanism. In a general case, much more effort is required to recover a distribution of interest.

\subsection{\texorpdfstring{Assuming G\textsubscript{alt2}}{Assuming Galt2}} \label{subsec:ident_Galt2}

We consider a second plausible alternative m-DAG, $\mathbf{G}_{alt2}$, which is equivalent to $\mathbf{G}_{main}$ in Figure \ref{fig:full_m-DAG}, with the addition of a direct effect of $BHV$ on $MEMS$ for weeks $6$, $36$, $48$, $60$ and $84$. In this model, we propose that behavioural pattern may affect adherence. \\

Under $\mathbf{G}_{alt2}$, the identified expression for the causal effect of interest, $\theta_{36}$, is as follows:
{\small{
\begin{flalign} \label{eq:ident_theta36_alt2}
\begin{split}
\theta_{36} = \sum_{\substack{Age,Sex,CoMo_0,Weight_0, \\ VL_0,CoMo_6,VL_6}}
& P(VL_{36}|Age,Sex,BHV,CoMo_0,Weight_0,Genotype,MEMS_6, \\
& Weight_6, Dose_6, EFV_6 = a,VL_0, CoMo_6, VL_6, MEMS_{36}, Weight_{36}, \\
& Dose_{36}, EFV_{36} = a) \\
& P(VL_6|Age,Sex,CoMo_0,Weight_0,Genotype,MEMS_6, \\
& Weight_6,Dose_6,EFV_6 = a,VL_0) \\
& P(CoMo_6|Age,Sex,CoMo_0,Weight_0,Genotype,VL_0) \\
& P(Weight_0|Age,Sex)P(CoMo_0|Age,Sex) P(VL_0) P(Sex)P(Age).
\end{split}
\end{flalign}
}}
Note that even if the identified expression looks almost identical to the one we obtain under $\mathbf{G}_{main}$ or $\mathbf{G}_{alt1}$ (compare Equation \ref{eq:ident_theta36_IDC}), the recoverability result may differ due to the different conditional independence statements that hold in $\mathbf{G}_{alt2}$, assuming it is faithful.

\section{Simulation Studies\label{app:simulation}}%

\subsection{DGP for Simulations 1 and 2 \label{app:simulation_12}}

Both baseline data ($t=0$) and follow-up data ($t=1,\ldots,5$) were created using structural equations with the $R$-package \texttt{simcausal}. Note that the follow-up time points $1$ through $5$ correspond to the study weeks $6$, $36$, $48$, $60$ and $84$. The distributions listed below, in temporal order, describe the data-generating process. Our baseline data consists of $Sex$, $Genotype$, $\log(\text{age})$ ($Age$), $\log(\text{weight})$ ($Weight$), the respective Nucleoside Reverse Transcriptase Inhibitor ($NRTI$), and a proxy for socio-economic status (SES). The time-varying variables are co-morbidities (CoMo), efavirenz dose (Dose), efavirenz mid-dose concentration (EFV), elevated viral load (= viral failure, VL), adherence (measured through memory caps, MEMS), missed visit (MV), technical issues (TI), and the missingness indicators for $EFV$, $Weight$, $VL$, and $MEMS$, respectively.
In addition to Bernoulli ($B$), Poisson ($Poisson$), Multinominal ($MN$) and Normal ($N$) distributions, we also use truncated normal distributions, denoted by $N_{[a,a_1,a_2,b,b_1,b_2]}$, where $a$ and $b$ are the truncation levels. Values smaller than $a$ are replaced by a random draw from a $U(a_1,a_2)$ distribution and values greater than $b$ are drawn from a $U(b_1,b_2)$ distribution, where $U$ refers to a continuous uniform distribution. For the specified multinomial distributions, probabilities are normalized, if required, to ensure they add up to 1. \\

The DGP corresponding to $\mathbf{G}_{main}$ in Figure \ref{fig:full_m-DAG} is as follows:

\vspace*{0.5cm}

{\scriptsize{
\noindent For $t=0$:
\noindent
\begin{eqnarray*}
\text{Sex}_0 &\sim& \text{B}(p=0.5) \\
\text{Genotype}_0 &\sim& \text{MN} \left( \begin{array}{ll}
               p1= 1/( 1+ \exp(-(-0.103 + \text{I}(\text{Sex}_0 = 1) \times 0.223 + \text{I}(\text{Sex}_0 = 0) \times 0.173)))\,,\\
               p2= 1/( 1+ \exp(-(-0.086 +  \text{I}(\text{Sex}_0 = 1) \times 0.198 + \text{I}(\text{Sex}_0 = 0) \times 0.214)))\,,\\
               p3= 1/( 1+ \exp(-(-0.309+  \text{I}(\text{Sex}_0 = 1) \times 0.082 +  \text{I}(\text{Sex}_0 = 0) \times  0.170 )))\\
               \end{array}
               \right) \\
\text{Age}_0 &\sim & N_{[0.693, 0.693, 1, 2.8, 2.7, 2.8]}( \mu = 1.501, \sigma = 0.369) \\
\text{Weight}_0 &\sim & N_{[2.26, 2.26, 2.67, 3.37, 3.02, 3.37]}( \mu = (1.5 + 0.2 \times \text{Sex}_0 + 0.774 \times \text{Age}_0) \times 0.94), \sigma = 0.369) \\
\text{NRTI}_0 &\sim& \text{MN} \left(  \begin{array}{ll}
               p1= 1/( 1+ \exp(-(-0.006 + \text{I}(\text{Age}_0>1.4563) \times  \text{Age}_0 \times 0.1735 +
               \text{I}(\text{Age}_0 \leq 1.4563) \times \text{Age}_0 \times 0.1570)))\,,\\
               p2= 1/( 1+ \exp(-(-0.006  + \text{I}(\text{Age}_0 > 1.4563) \times  \text{Age}_0  \times 0.1735 +
               \text{I}(\text{Age}_0 \leq 1.4563) \times \text{Age}_0 \times 0.1570)))\,,\\
               p3= 1/( 1+ \exp(-(-0.006  + \text{I}(\text{Age}_0 > 1.4563) \times  \text{Age}_0  \times 0.1570 +
               \text{I}(\text{Age}_0 \leq .14563) \times \text{Age}_0 \times 0.1818))) \\
               \end{array}
               \right) \\
\text{CoMo}_0 & \sim & \text{B}(p=0.15)\\
\text{VL}_0 & \sim & \text{B}(p= 1 - (1/(1+\exp(-(0.4+1.9 \times \sqrt{\text{EFV}_0})))))\\
\text{SES}_0 & \sim & Poisson(\lambda = 3)\\
\text{MV}_0 & \sim & \text{B}(p = 1/(1+\exp(-(-2.95 + 0.1 \times \text{SES}_0))))\\
\end{eqnarray*}

\noindent For $t=1$:
\noindent
\begin{eqnarray*}
\text{Dose}_1 & \sim &\text{MN} \left( \begin{array}{ll}
              p1= 1/( 1+ \exp(-(5 + \sqrt{(\text{Weight}_1)} \times 8 -  \text{Age}_0 \times 10)))\,, \\
              p2= 1/( 1+ \exp(-(4 + \sqrt{(\text{Weight}_1)} \times 8.768 - \text{Age}_0 \times 9.06))) \,, \\
              p3= 1/( 1+ \exp(-(3 + \sqrt{(\text{Weight}_1)} \times 6.562 - \text{Age}_0 \times 8.325)))\,,\\
              p4= 1-(p1+p2+p3) \\
               \end{array}
               \right)
\end{eqnarray*}

\noindent For $t\geq0$:
\noindent
\begin{eqnarray*}
\text{TI}_t & \sim & \text{B}(p = 0.05))))\\
\text{M}_{EFV_t} & \sim & \text{B}(p = I(\text{MV}_t=1) + I(\text{MV}_t=0) \times I(\text{TI}_t=1) \times 0.5))))\\
\text{M}_{Weight_t} & \sim & \text{B}(p = I(\text{MV}_t=1))\\
\text{M}_{VL_t} & \sim & \text{B}(p = I(\text{MV}_t=1) + I(\text{MV}_t=0) \times I(\text{TI}_t=1) \times 0.5))))\\
\end{eqnarray*}

\noindent For $t\geq1$:
\noindent
\begin{eqnarray*}
\text{MEMS}_t &\sim& \text{B}(p= 1/(1+ \exp(-(0.71 + \text{CoMo}_{t-1} \times 0.31 + \text{MEMS}_{t-1} \times \text{I}(t\geq2) \times 0.31)))), \quad [\text{assume}  \ \text{MEMS}_0 = 0] \\
\text{Weight}_t &\sim& N_{[2.26,2.26,2.473,3.37,3.2,3.37]}(\mu = \text{Weight}_{t-1}\times 1.04 -0.05 \times \text{I}(\text{CoMo}_{t-1}=1),\, \sigma=0.4 )  \\
\text{CoMo}_t &\sim & \text{B}(p = 1 - (1/( 1+ \exp(- (0.5 \times \text{I}(\text{CoMo}_{t-1}=1) + \text{Age}_{0} \times 0.1 + \text{Weight}_{t-1} \times 0.1))))) \\
\text{EFV}_t &\sim & N_{[0.2032, 0.2032, 0.88, 21.84, 8.37, 21.84]}(\mu = 0.1 \times \text{Dose}_t + 0.1 \times \text{MEMS}_{t} + \text{I}(\text{Genotype}_0 \leq 2) \times 2.66 \\&& \quad +  \text{I}(\text{Genotype}_0 = 3) \times 4.6,  \sigma = 4.06)\\
\text{VL}_t & \sim & \text{B}(p= 1- (1/(1+\exp(-(1-0.6\times\text{I}(t=1)-1.2\times\text{I}(t=4) + 0.1 \times \text{CoMo}_{t-1} + (2 - 0.2\times\text{I}(t=3)) \times \sqrt{\text{EFV}_t}))))) \\
\text{MV}_t & \sim & \text{B}(p = 1/(1+\exp(-(-2.95 + 0.1 \times \text{SES}_0 + \text{MV}_{t-1}))))\\
\text{M}_{MEMS_t} & \sim & \text{B}(p = 1/(1+\exp(-(0.5 \times I(\text{TI}_t = 1) + 0.2))))\\
\end{eqnarray*}

\noindent For $t\geq2$:
\noindent
\begin{eqnarray*}
\text{Dose}_t & \sim & \text{MN} \left(\begin{array}{ll}
               p1 = (1/( 1+ \exp(-(4 + \text{Dose}_{t-1} \times 0.5 + \sqrt{\text{Weight}_t} \times 4  -  \text{Age}_0 \times 10)))\,,\\
               p2 = (1/( 1+ \exp(-(-8 + \text{Dose}_{t-1} \times 0.5  + \sqrt{\text{Weight}_t} \times 8.568  -  \text{Age}_0 \times 9.06)))\,,\\
               p3 = (1/( 1+ \exp(-(20 + \text{Dose}_{t-1} \times 0.5 + \sqrt{\text{Weight}_t} \times 6.562  -  \text{Age}_0 \times 18.325)))\,,\\
               p4 = 1-(p1+p2+p3) \\
               \end{array}
               \right) \\
\end{eqnarray*}
}}
The DGP for $\mathbf{G}_{alt1}$ (including the blue dashed lines in Figure \ref{fig:full_m-DAG}) coincides with the DGP above, except for the structural equations for $MV_t$, $t \in \{0, 6, 36, 48, 60, 84\}$. These are specified as follows for $\mathbf{G}_{alt1}$: \\

{\scriptsize{
\noindent For $t=0$:
\noindent
\begin{eqnarray*}
\text{MV}_0 & \sim & \text{B}(p = 1/(1+\exp(-(-2.95 + 0.1 \times \text{SES}_0 + 2 \times \text{VL}_0))))\\
\end{eqnarray*}
\noindent For $t\geq1$:
\noindent
\begin{eqnarray*}
\text{MV}_t & \sim & \text{B}(p = 1/(1+\exp(-(-2.95 + 0.1 \times \text{SES}_0 + \text{MV}_{t-1} + 2 \times \text{VL}_t))))\\
\end{eqnarray*}
}}
Thus, $MV_t$ additionally depends on $VL_t$, $t \in \{0, 6, 36, 48, 60, 84\}$, which corresponds to the dashed blue lines in the DAG. \\

The DGP for $\mathbf{G}_{alt2}$ (where $SES$ is a cause of $MEMS_t$, $t \in \{6, 36, 48, 60, 84\}$) coincides with the DGP for $\mathbf{G}_{main}$ above, except for the structural equations for $MEMS_t$, $t \in \{6, 36, 48, 60, 84\}$. These are specified for $\mathbf{G}_{alt2}$ as follows: \\
{\scriptsize{
\noindent For $t\geq1$:
\noindent
\begin{eqnarray*}
\text{MEMS}_t \sim \text{B}(p= 1/(1+ \exp(-(0.71 + \text{CoMo}_{t-1} \times 0.31 + \text{MEMS}_{t-1} \times \text{I}(t\geq2) \times 0.31 - \text{SES}_{0} \times 0.5))))
\end{eqnarray*}
}}

After generating the data set using the structural equations, we introduce missing values based on the missingness indicators: if a missingness indicator equals 1, the corresponding covariate value is set to \texttt{NA}. In \emph{Simulation 1}, we ignore the missingness indicators for $EFV_t$, $t \in \{6, 36, 48, 60, 84\}$, and $Weight_t$, $t \in \{0, 6, 36, 48, 60, 84\}$, and generate missingness only in $VL_t$ and $MEMS_t$, $t \in \{0, 6, 36, 48, 60, 84\}$. In \emph{Simulation 2}, missingness is introduced in $EFV_t$, $t \in \{6, 36, 48, 60, 84\}$, $Weight_t$, $VL_t$ and $MEMS_t$, $t \in \{0, 6, 36, 48, 60, 84\}$.

\subsection{Results\label{app:simulation2}}
The results below are based on Simulation 2 as defined in Section \ref{sec5}.

\begin{figure}[!ht]
    \centering
    \subfloat[Data simulated under the main DAG $\mathbf{G}_{\text{main}}$]{
        \includegraphics[width=0.48\textwidth,height=0.8\textheight,keepaspectratio]{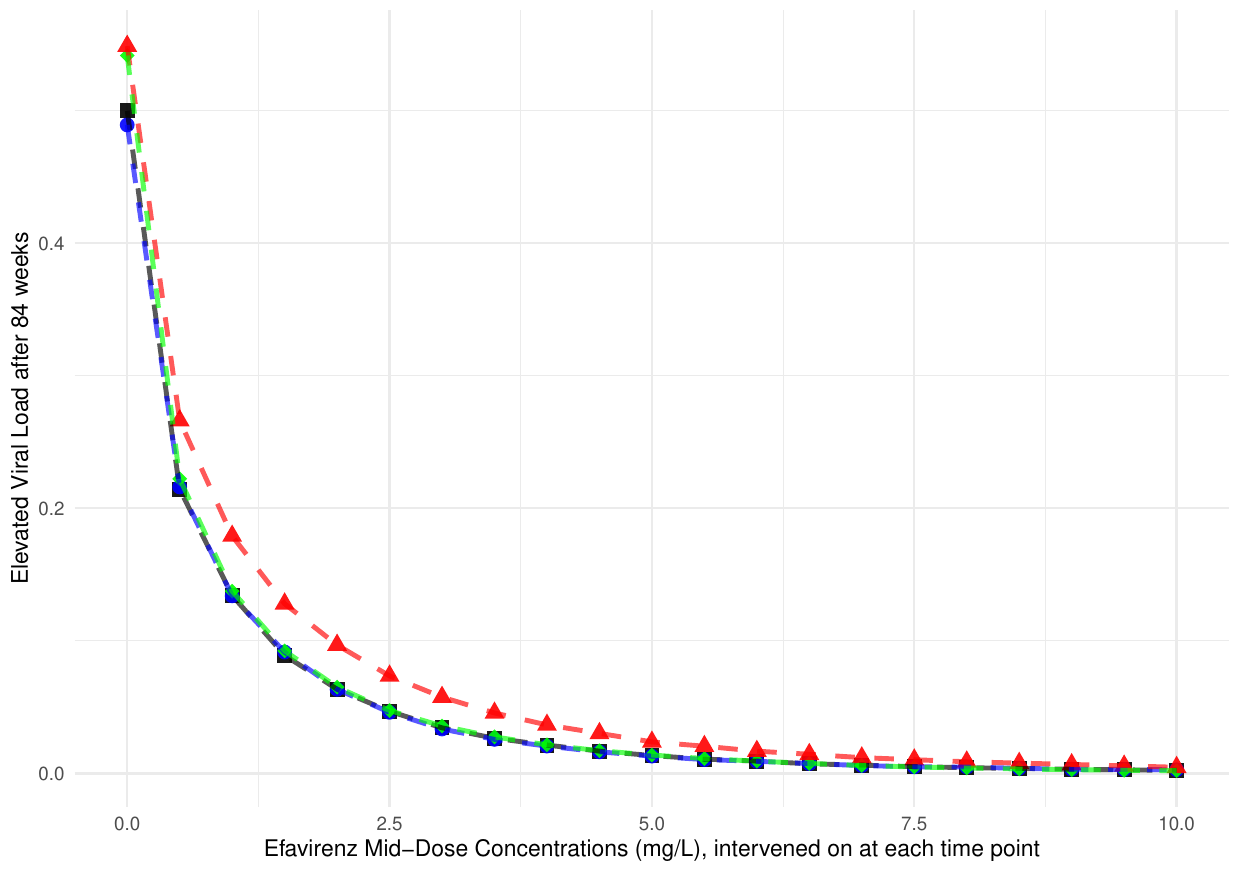}
        \label{fig:CCRC_vl84_main_4miss}
    }\hfill
    \subfloat[Data simulated under the alternative DAG $\mathbf{G}_{\text{alt1}}$]{
        \includegraphics[width=0.48\textwidth,height=0.8\textheight,keepaspectratio]{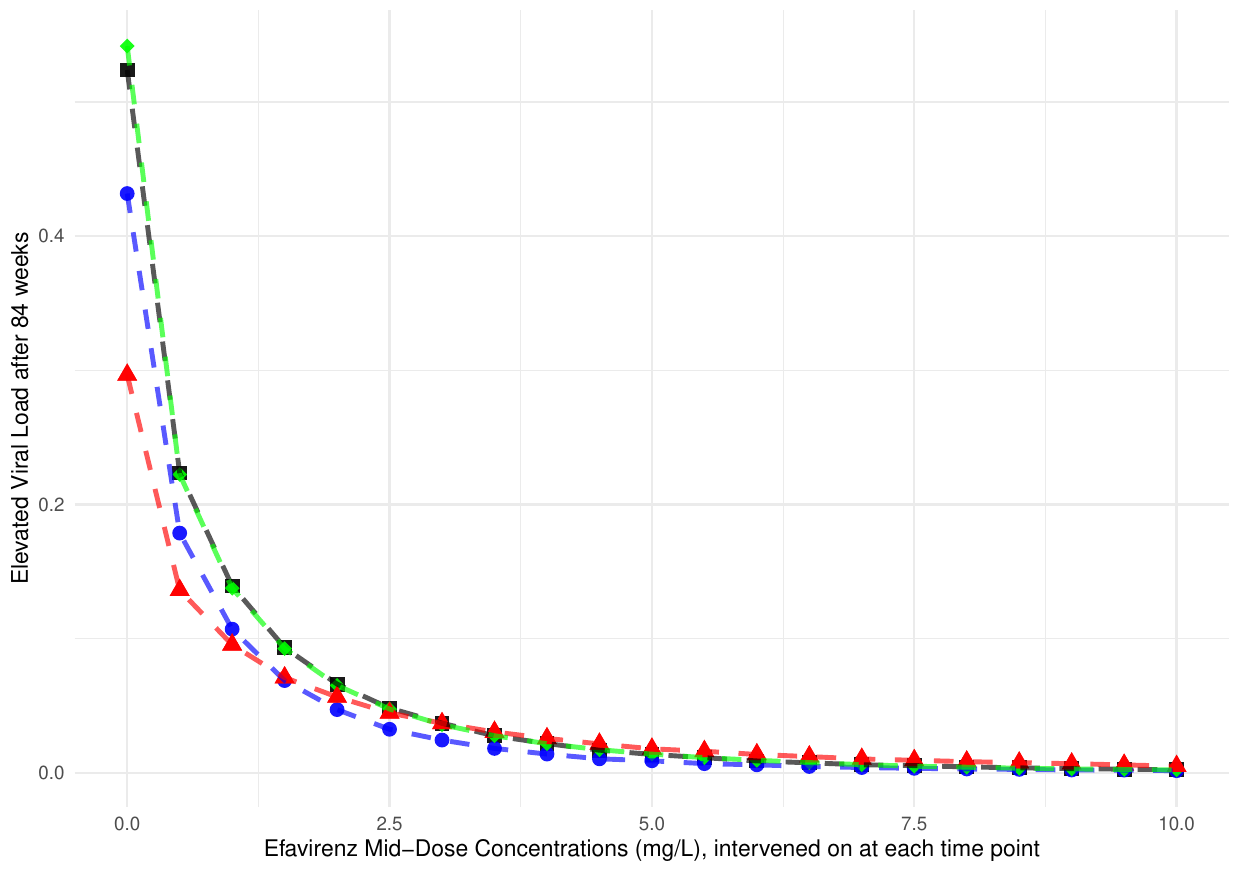}
        \label{fig:CCRC_vl84_altern_4miss}
    }
    \caption{Estimated CCRCs for the probability of viral failure after 84 weeks. (a) Data simulated under the main DAG $\mathbf{G}_{\text{main}}$. (b) Data simulated under the alternative DAG $\mathbf{G}_{\text{alt1}}$. Causal effects were estimated on complete data (black squares), incomplete data using available cases (blue dots), incomplete data using multiple imputation (red triangles) and counterfactual data (green diamonds, true CCRC); results represent the mean over $1000$ seeds.}
     \label{fig:CCRC_vl84_4miss}
\end{figure}

The following results are based on simulation 1 as defined in Section \ref{sec5} under the assumption of $\mathbf{G}_{alt2}$ being the true underlying causal m-DAG.

\begin{figure}[!ht]
    \begin{minipage}{0.48\textwidth}
        \centering
        \includegraphics[width=1.0\textwidth,height=0.8\textheight,keepaspectratio]{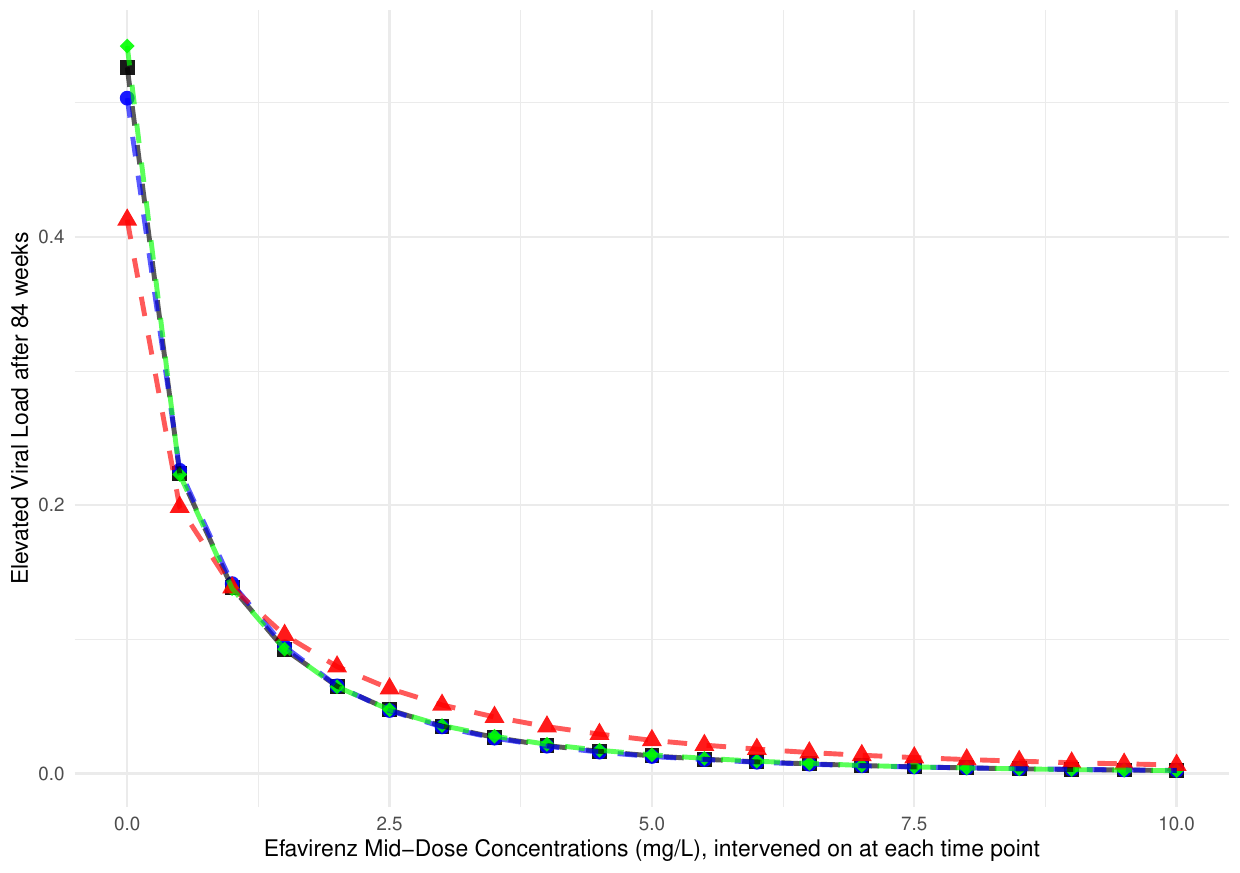}
        \caption{Estimated CCRCs for the probability of viral failure after 84 weeks based on data simulated under the main DAG $\mathbf{G}_{alt2}$. Causal effects were estimated on complete data (black squares), incomplete data using available cases (blue dots), incomplete data using multiple imputation (red triangles)
        and counterfactual data (green diamonds, true CCRC); results represent the mean over $1000$ seeds.}
        \label{fig:CCRC_vl84_altern2_2miss}
    \end{minipage}
    \hfill
    \begin{minipage}{0.48\textwidth}
        \centering

    \end{minipage}
\end{figure}

\subsection{Monte Carlo Confidence Intervals\label{app:mc_ci}}

The Monte Carlo confidence intervals in Tables \ref{tab:mi_main}, \ref{tab:ac_alt1} and \ref{tab:mi_alt1} are reported to determine whether the differences between MI estimates and true values (main and alternative m-DAGs, $\mathbf{G}_{main}$ and $\mathbf{G}_{alt1}$), and between available case analysis estimates and true causal effects $\theta_{84}$ (alternative m-DAG $\mathbf{G}_{alt1}$) stem from simulation uncertainty or indicate systematic deviations. A 95\% confidence interval [mean difference $- 2 \cdot SE$; mean difference $+ 2 \cdot SE$], with the standard error $SE$ computed as $sd(estimate)/\sqrt{(\#runs)}$, excluding zero suggests that the differences can likely not be explained by simulation uncertainty, indicating a bias caused by the estimation approach.

\begin{table}[!ht]
    \centering
    \caption{Monte Carlo confidence intervals for the difference between estimates and true causal effects across efavirenz (EFV) concentrations. 
    (a) Multiple imputation estimates under the main m-DAG $\mathbf{G}_{\text{main}}$. 
    (b) Available case estimates under the alternative m-DAG $\mathbf{G}_{\text{alt1}}$. 
    (c) Multiple imputation (MI) estimates under the alternative m-DAG $\mathbf{G}_{\text{alt1}}$.}
    \label{tab:monte_carlo}
    \subfloat[Multiple imputation under $\mathbf{G}_{\text{main}}$]{
        \begin{minipage}{0.31\linewidth}
            \centering
            \begin{tabular}{rrr}
                \toprule
                EFV & Lower & Upper \\
                \midrule
                0.0 & -0.19441 & -0.19351 \\
                0.5 & -0.04522 & -0.04452 \\
                1.0 & -0.00881 & -0.00824 \\
                1.5 & 0.00696 & 0.00749 \\
                2.0 & 0.01493 & 0.01539 \\
                2.5 & 0.01774 & 0.01817 \\
                3.0 & 0.01842 & 0.01880 \\
                3.5 & 0.01812 & 0.01847 \\
                4.0 & 0.01721 & 0.01754 \\
                4.5 & 0.01626 & 0.01656 \\
                5.0 & 0.01502 & 0.01530 \\
                5.5 & 0.01403 & 0.01429 \\
                6.0 & 0.01309 & 0.01335 \\
                6.5 & 0.01191 & 0.01214 \\
                7.0 & 0.01123 & 0.01145 \\
                7.5 & 0.01051 & 0.01071 \\
                8.0 & 0.00952 & 0.00971 \\
                8.5 & 0.00886 & 0.00904 \\
                9.0 & 0.00810 & 0.00827 \\
                9.5 & 0.00750 & 0.00766 \\
                10.0 & 0.00685 & 0.00701 \\
                \bottomrule
            \end{tabular}
        \end{minipage}
        \label{tab:mi_main}
    }\hfill 
    \subfloat[Available case under $\mathbf{G}_{\text{alt1}}$]{
        \begin{minipage}{0.31\linewidth}
            \centering
            \begin{tabular}{rrr}
                \toprule
                EFV & Lower & Upper \\
                \midrule
                0.0 & -0.12787 & -0.10112 \\
                0.5 & -0.04754 & -0.03497 \\
                1.0 & -0.03208 & -0.02512 \\
                1.5 & -0.02465 & -0.02075 \\
                2.0 & -0.01873 & -0.01642 \\
                2.5 & -0.01470 & -0.01320 \\
                3.0 & -0.01195 & -0.01092 \\
                3.5 & -0.01012 & -0.00935 \\
                4.0 & -0.00823 & -0.00759 \\
                4.5 & -0.00677 & -0.00620 \\
                5.0 & -0.00558 & -0.00509 \\
                5.5 & -0.00450 & -0.00405 \\
                6.0 & -0.00358 & -0.00317 \\
                6.5 & -0.00296 & -0.00257 \\
                7.0 & -0.00222 & -0.00186 \\
                7.5 & -0.00175 & -0.00141 \\
                8.0 & -0.00145 & -0.00115 \\
                8.5 & -0.00114 & -0.00085 \\
                9.0 & -0.00088 & -0.00062 \\
                9.5 & -0.00074 & -0.00048 \\
                10.0 & -0.00058 & -0.00034 \\
                \bottomrule
            \end{tabular}
        \end{minipage}
        \label{tab:ac_alt1}
    }\hfill 
    \subfloat[Multiple imputation under $\mathbf{G}_{\text{alt1}}$]{
        \begin{minipage}{0.31\linewidth}
            \centering
            \begin{tabular}{rrr}
                \toprule
                EFV & Lower & Upper \\
                \midrule
                0.0 & -0.29980 & -0.29904 \\
                0.5 & -0.10118 & -0.10062 \\
                1.0 & -0.04923 & -0.04876 \\
                1.5 & -0.02354 & -0.02312 \\
                2.0 & -0.00888 & -0.00849 \\
                2.5 & -0.00111 & -0.00076 \\
                3.0 & 0.00326 & 0.00358 \\
                3.5 & 0.00583 & 0.00614 \\
                4.0 & 0.00727 & 0.00755 \\
                4.5 & 0.00804 & 0.00830 \\
                5.0 & 0.00832 & 0.00857 \\
                5.5 & 0.00849 & 0.00871 \\
                6.0 & 0.00851 & 0.00872 \\
                6.5 & 0.00815 & 0.00835 \\
                7.0 & 0.00819 & 0.00838 \\
                7.5 & 0.00795 & 0.00814 \\
                8.0 & 0.00749 & 0.00766 \\
                8.5 & 0.00718 & 0.00734 \\
                9.0 & 0.00675 & 0.00691 \\
                9.5 & 0.00640 & 0.00655 \\
                10.0 & 0.00605 & 0.00620 \\
                \bottomrule
            \end{tabular}
        \end{minipage}
        \label{tab:mi_alt1}
    }
\end{table}

\clearpage

\section*{Competing interests}
No competing interest is declared.

\section*{Author contributions statement}

The study was designed by MS, PD and HM. Data was collected and interpreted by PD, HM. All authors reviewed the study design and interpreted the data. Methods development was lead by AH. The first draft of the article was written by AH. All authors reviewed this and the following drafts, revising it critically for its content. All authors have read and approved the final version and agree with the manuscript’s conclusion.

\section*{Acknowledgments}
We are grateful for the support of the CHAPAS-3 trial team, their advice regarding the data analysis and making their data available to us. We would like to thank David Burger, Sarah Walker, Di Gibb and Andrzej Bienczak for their help in interpreting the data. We thank Elizabeth Kaudha and Victor Musiime for discussing reasons for missingness with us. We would further like to acknowledge the help of Alexander Szubert in constructing some of the variables.  Michael Schomaker is supported by the German Research Foundations (DFG) Heisenberg Programm (grants 465412241 and 465412441). The CHAPAS-3 trial was funded by the European Developing Countries Clinical Trials Partnership (IP.2007.33011.006), Medical Research Council UK (MC\_UU\_00004/03), Department for International Development UK, and Ministerio de Sanidady Consumo Spain. Cipla Ltd donated first-line antiretrovirals. We thank all the children, carers, and staff from all the centres participating in the CHAPAS-3 trial.

\section*{Data Availability}

The code used for conducting the simulation study in Section \ref{sec5} is accessible at the following GitHub repository: \url{https://github.com/aholovchak/Recoverability-of-causal-effects}.

\end{document}